\documentclass[twocolumn,showpacs,prl]{revtex4-2}
\usepackage{graphicx,amsmath,mathrsfs}
\usepackage{amssymb}
\usepackage{amsthm,multirow}
\usepackage{bm,bbm}
\usepackage{subfigure,xcolor}   
\usepackage{color}

\def\bc{\begin{center}}
\def\ec{\end{center}}

\usepackage{physics}
\usepackage{dsfont}
\usepackage[unicode=true,pdfusetitle,
bookmarks=true,bookmarksnumbered=false,bookmarksopen=false,
breaklinks=true,pdfborder={0 0 0},pdfborderstyle={},backref=false,colorlinks=false]{hyperref}
\hypersetup{linkcolor=blue,citecolor=blue,urlcolor=blue}

\renewcommand\vec{\mathbf}		                             
\newcommand{\unitvec}[1]{\hat{\mathbf{#1}}}			
\newcommand{\sgn}{\mathrm{sgn}}

\newcommand{\HH}{{\cal H}}

\graphicspath{{./}{./figs/}}

\begin{document}
\title{
SU($N$) altermagnetism: Lattice models, magnon modes, and flavor-split bands
}
\author{Pedro M. C\^onsoli}
\author{Matthias Vojta}
\affiliation{Institut f\"ur Theoretische Physik and W\"urzburg-Dresden Cluster of Excellence ct.qmat, Technische Universit\"at Dresden, 01062 Dresden, Germany}

\begin{abstract}
Altermagnetism, a type of magnetic order that combines properties of ferro- and antiferromagnets, has stirred great interest lately not only as a promising source of spintronics applications, but also as a potential gateway to exotic phases of matter. Here, we demonstrate how to generalize collinear altermagnetism to SU($N$) magnets with $N>2$. Guided by symmetry principles, we present a recipe to construct Heisenberg models for such generalized altermagnets and apply it explicitly for $N=3,4$. Using flavor-wave theory, we compute the excitation spectrum of a two-dimensional SU(3) model and show that it exhibits magnon bands with altermagnetic splitting according to magnetic quantum numbers; we connect this quantum-number splitting to the frequently used concept of magnon chirality. We also compute the electronic band structure for a metallic system of the same symmetry and map out the polarization of the resulting flavor-split bands.
\end{abstract}

\date{\today}

\maketitle


Altermagnetism is an unconventional type of magnetic order, beyond standard ferro- or antiferromagnetism, which is currently attracting significant attention \cite{smejkal22,smejkal_rev}. Despite having no uniform magnetization, altermagnets display a variety of properties that are typically associated with ferromagnets, such as electronic bands that are spin-split even in the absence of spin-orbit coupling \cite{hayami19,hayami20} and associated macroscopic magnetotransport phenomena \cite{smejkal_rev2}. These properties render altermagnetic states interesting both on fundamental \cite{chakraborty24,sim24,sobral24,mcclarty24} and applied \cite{smejkal_rev,gonzalez21} grounds. Microscopically, the magnetic sublattices of an altermagnet are related by symmetry operations other than translation and inversion, giving rise to spin polarization in reciprocal space. A formal description based on a spin-space symmetry classification was given in Ref.~\onlinecite{smejkal22}. In addition to special conduction-band properties, altermagnets also feature a nontrivial magnon spectrum: the dispersions of magnons of different chiralities split in an anisotropic fashion, again dictated by the underlying spin-space symmetries \cite{smejkal23}.
The vast majority altermagnetic states discussed to date feature collinear magnetic order of SU(2) moments, including the experimental realizations in RuO$_2$ \cite{feng22,tschirner23}, MnTe \cite{betancourt23}, and thin-film Mn$_5$Si$_3$ \cite{reichlova20}. Extensions to noncollinear spin configurations have been discussed in Refs.~\onlinecite{cheong20,cheong24}, as well as in the context of $p$-wave magnets \cite{hellenes24}.

The aim of this paper is to generalize the concept of altermagnetism to higher SU($N$) symmetry groups. For concreteness, we focus primarily on SU(3) and construct two models on a modified triangular lattice with a three-color ground state whose magnetic sublattices are related by rotation and mirror symmetries, but not by translations nor inversions. We analyze the first, Heisenberg-type model via flavor-wave theory and show that its magnon spectrum features nondegenerate bands, each associated with a unique set of magnetic quantum numbers. We also elucidate that these bands emerge in pairs of opposite chirality. Next, by including charge carriers, we obtain an SU(3) double-exchange model. Upon computing its electronic band structure, we find that the bands are flavor-split except along high-symmetry lines in reciprocal space, where (partial) degeneracies are imposed by spin-point-group symmetries, i.e., combinations of lattice point-group operations and independent transformations in spin space \cite{litvin74,litvin77,smejkal22}.
For comparison, we perform the same steps for models with collinear SU(2) spin order on a checkerboard lattice, which has recently appeared as a motif for altermagnetism \cite{knap23}. In doing so, we demonstrate that our approach reproduces multiple known properties of SU(2) altermagnets. We conclude by discussing applications of our ideas.


\begin{figure}[!tb]
\includegraphics[width=\columnwidth]{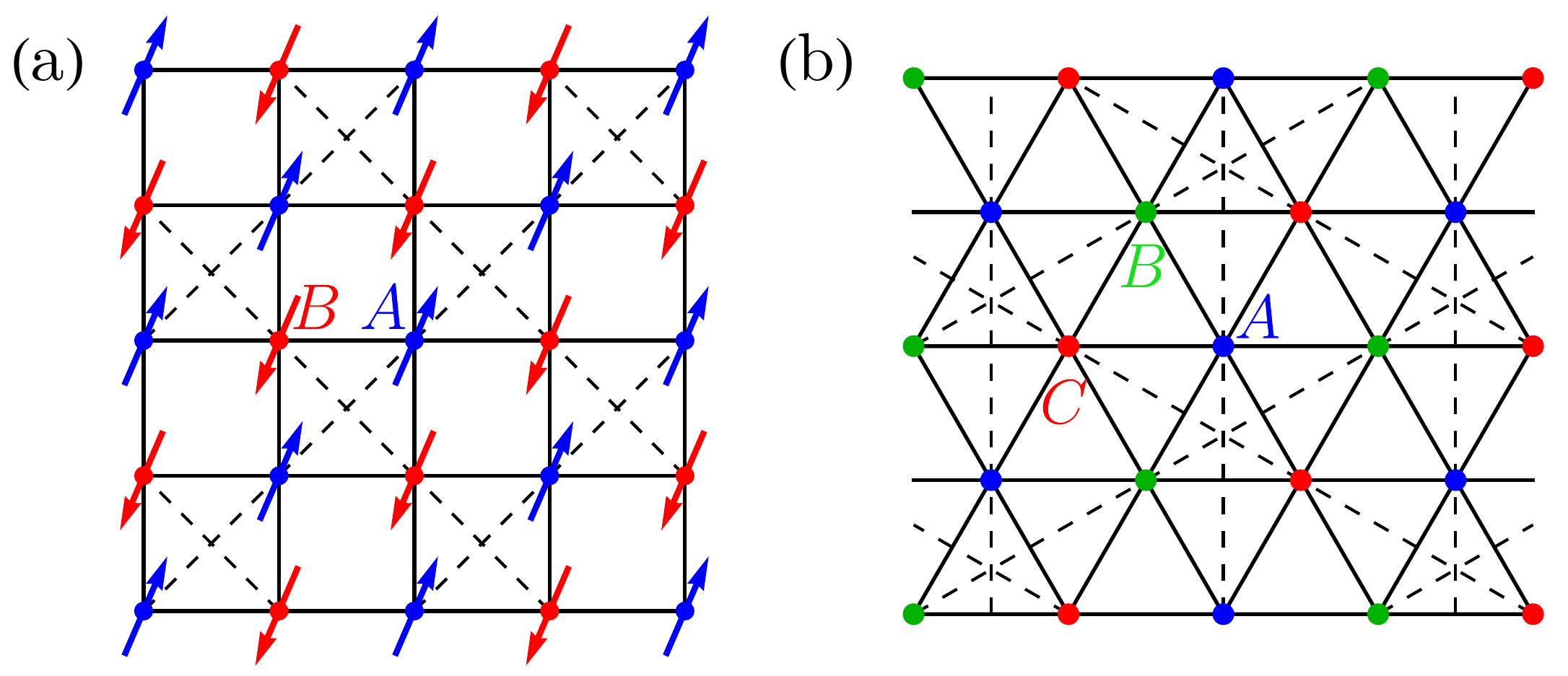}
\caption{Lattice structures of altermagnetic Heisenberg models:
(a) Checkerboard lattice with collinear SU(2) order;
(b) Hexatriangular lattice with three-color SU(3) order.
Solid (dashed) lines show couplings $J_1$ ($J_2$).}
\label{fig:latt1}
\end{figure}

\paragraph{Heisenberg models of altermagnets. --}
Despite resembling antiferromagnets in that they have no net magnetization, altermagnets lack the translation and inversion symmetries that would otherwise map different magnetic sublattices onto each other. In real-world materials, this symmetry breaking is often caused by the presence of non-magnetic atoms that lower the symmetry of the lattice of magnetic atoms. Here, however, we will implement it on the level of simple models. We will first construct Heisenberg-type spin models for insulating altermagnets and add conduction electrons later on. We thus start by considering SU($N$)-symmetric Hamiltonians \cite{affleck85,papa88,read89,joshi99,bauer12}
\begin{equation}
	\label{eq:heis}
	\mathcal{H}_S = \frac{1}{2N} \sum_{ij} \sum_{\alpha,\beta=1}^{N} J_{ij} S_{i}^{\alpha\beta} S_{j}^{\beta\alpha},
\end{equation}
where $\alpha,\beta$ are flavor indices and $J_{ij}$ is the exchange coupling between sites $i$ and $j$. For every $j$, the $S_{j}^{\alpha\beta}$ form a set of SU($N$) generators that obey the commutation relations $\left[ S_{j}^{\alpha\beta}, S_{j}^{\mu\nu} \right] = \delta_{\beta\mu} S_{j}^{\alpha\nu} - \delta_{\nu\alpha} S_{j}^{\mu\beta}$. We will henceforth assume that the latter are in the $\left\{ M,0 \right\}$ totally symmetric irreducible representation (irrep) of SU($N$), corresponding to a Young tableau with $M$ boxes arranged in a single row. The local Hilbert space of a spin is then spanned by the eigenvectors shared between the $N$ diagonal generators $S^{\mu\mu}_j$, which satisfy the constraint $\sum_{\mu} S^{\mu\mu}_j = M \mathds{1}$ and have integer eigenvalues between $0$ and $M$. For $N=2$, the nondiagonal generators $S_{j}^{\alpha\beta}$, with $\alpha\ne\beta$, are nothing but the ladder operators $S_{j}^{\pm} = S_{j}^{x} \pm i S_{j}^{y}$ whereas $S_j^z = (S_j^{11} - S_j^{22})/2$.  Equation~\eqref{eq:heis} thus becomes $\mathcal{H}_{S} = \frac{1}{2}\sum_{ij} J_{ij} \vec{S}_i \cdot \vec{S}_j$, with spins $S=M/2$, apart from an irrelevant constant. In what follows, we will investigate these models in the semiclassical regime of large $M$. The results obtained in this manner will remain qualitatively correct for small $M$ provided that the respective magnetic order occurs.

The construction of the altermagnetic models proceeds in two steps:
(i) Given a lattice that can be partitioned into $N$ equivalent sublattices $\mu = 1, \ldots, N$, we start with a simple Hamiltonian whose ground state in the classical limit $M\to \infty$ is the $N$-color product state
\begin{equation}
	\ket{\psi_0} = \bigotimes_{i} \bigotimes_{\mu = 1}^N \ket{\mu}_{i\mu}.
	\label{eq:Ncolorstate}
\end{equation}
Here, each site $(i\mu)$ is identified by its magnetic unit cell $i$ and sublattice $\mu$, whereas $\ket{\mu}$ denotes the state with highest weight with respect to $S^{\mu\mu}$, i.e., $S_{i\mu}^{\alpha\alpha}\ket{\mu}_{i\mu} = \delta_{\alpha\mu} M\ket{\mu}_{i\mu}$. For $N=2$, Eq.~\eqref{eq:Ncolorstate} is simply a collinear N\'eel state.
(ii) Augment the model by including couplings $J_{ij}$ that break any translation or inversion symmetry connecting  the different sublattices, while preserving $\ket{\psi_0}$ and the equivalence of the sublattices under rotations.

To connect to previous work, we first apply this procedure to the SU(2) case. For step (i), we consider spins $S$ interacting via nearest-neighbor antiferromagnetic couplings, $J_1$, on a square lattice. Collinear N\'eel order divides the system into sublattices $A\equiv 1$ and $B\equiv 2$. In step (ii), we add selected second-neighbor couplings, $J_2$, in the pattern of a checkerboard lattice, Fig.~\ref{fig:latt1}(a). This results in a two-site unit cell and $D_{4}$ point-group symmetry. Different magnetic sublattices are related either by mirror operations or by $90^\circ$ rotations about the center of an elementary square plaquette.

For the SU(3) case, we start step (i) with a nearest-neighbor model on the triangular lattice, where an antiferromagnetic coupling $J_1$ leads to three-sublattice, three-color order in the ground state for every $M\ge 1$ \cite{laeuchli06,tsunetsugu06,bauer12}. For step (ii), we add selected second-neighbor couplings $J_2$ as shown in Fig.~\ref{fig:latt1}(b). This results in a ``hexatriangular lattice'' with six types of elementary triangular plaquettes and a three-site unit cell. The magnetic sublattices are related by the $D_3$ point group, i.e., 120$^\circ$ rotations about the center of a filled triangle and reflections with respect to dashed lines in Fig.~\ref{fig:latt1}(b). Generalizations to higher dimensions and larger $N$ are clearly possible. In particular, a three-dimensional SU(4) model is presented in the Supplemental Material (SM) \cite{suppl}.

We will discuss the properties of these Heisenberg models at zero temperature and focus on the regime where $J_2/J_1$ is less than a certain critical ratio above which the semiclassical ordered state $\ket{\psi_0}$ changes.

\begin{figure}[!tb]
	\includegraphics[width=\columnwidth]{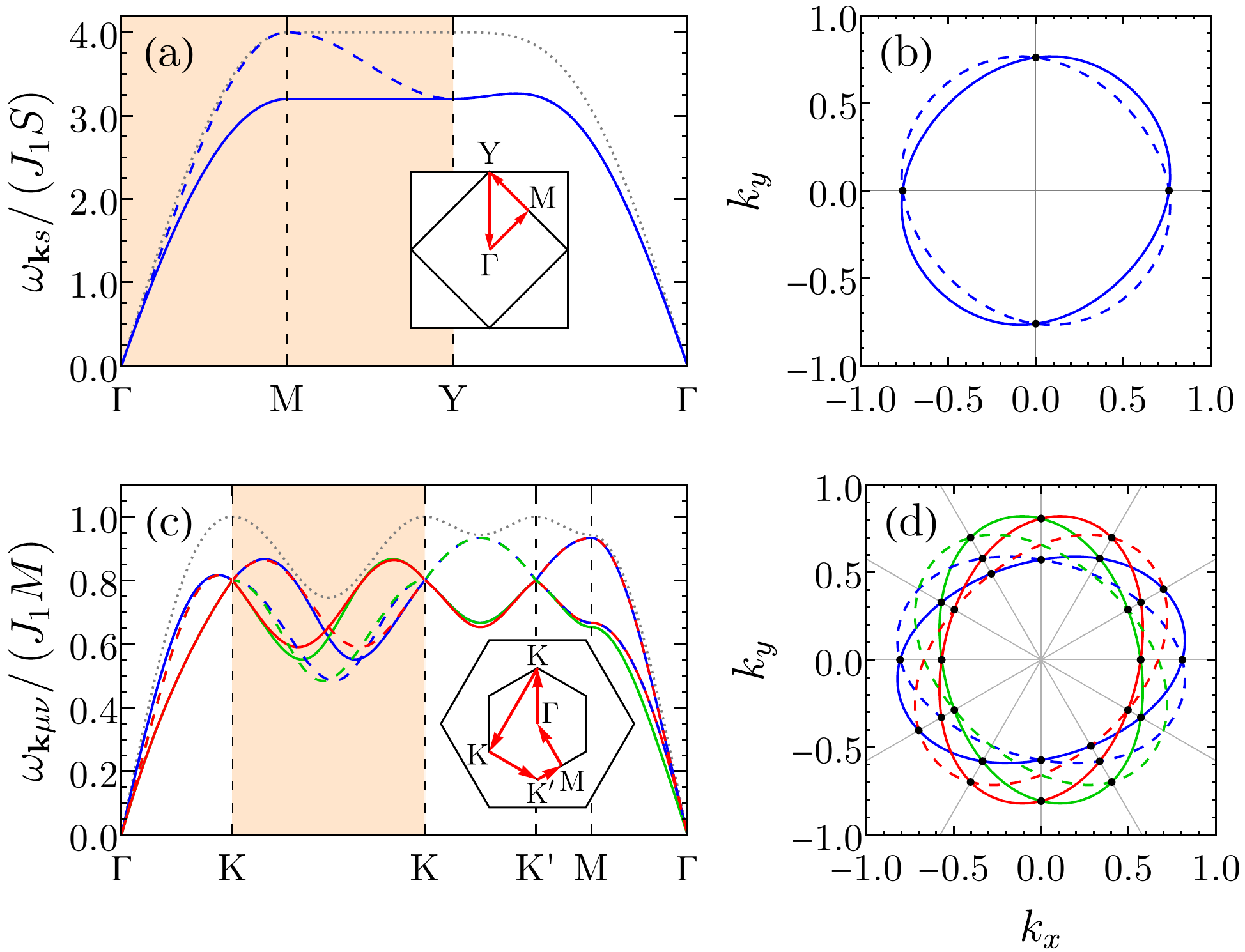}
	\caption{Magnon dispersions of the (a,b) SU(2) and (c,d) SU(3) Heisenberg altermagnets for $J_2/J_1 = 0.2$. Pairs of bands with the same color, but different line patterns, have opposite sets of magnetic quantum numbers and chiralities. In (c,d), the colors correspond to the sublattices that remain unaffected when a magnon of a given mode is created -- e.g., the bands $\omega_{\vec{k}BC}$ and $\omega_{\vec{k}CB}$ are blue [see Fig.~\ref{fig:latt1}(b)]. The left panels show the dispersions along paths depicted in the insets.  Sections where the complete band splitting is visible are highlighted in orange and the dispersion for $J_2 = 0$ is shown in a dotted gray line for comparison. The right panels depict isoenergy contours at (b) $E=1.8J_1$ and (d) $E=0.35J_1$.}
	\label{fig:spinw1}
\end{figure}

\paragraph{Spin waves and flavor waves. --}
In the SU(2) case, we performed standard linear spin-wave theory as in Ref.~\onlinecite{canals02}. Using a Holstein-Primakoff representation in terms of bosons $a_i$ and $b_i$, such that $S_{iA}^{z} = S - a_{i}^\dagger a_{i}$ and $S_{iB}^z = -S + b_{i}^\dagger b_{i}$, we find that the bilinear part of $\HH_S$ is
\begin{equation}
	\mathcal{H}_S^{(2)} = 4J_1S \sum_{\vec{k}}
	\Psi_\vec{k}^\dagger
	\begin{pmatrix}
		1-2\eta\Gamma_{\vec{k}1} & \gamma_{\vec{k}}\\
		\gamma_{\vec{k}} & 1-2\eta\Gamma_{\vec{k}2}
	\end{pmatrix}
	\Psi_\vec{k},
	\label{eq:H preBog SU2}
\end{equation}
where $\Psi_\vec{k}^\dagger = \begin{pmatrix} a_{\vec{k}}^\dagger & b_{-\vec{k}} \end{pmatrix}$, $\eta=J_{2}/\left(4J_1\right)$, and the sum runs over the first Brillouin zone. Furthermore, we have $\gamma_{\vec{k}} = \frac{1}{2} \left( \cos k_x + \cos k_y \right)$ and $\Gamma_{\vec{k}1,2} = 1 - \cos\left(k_x \pm k_y\right)$. The diagonalization of $\HH_S^{(2)}$ yields the spectrum \cite{canals02}
\begin{align}
	\frac{\omega_{\vec{k}\pm}}{4J_{1}S}
	= \sqrt{\left[ 1 - \eta \left(\Gamma_{\vec{k}1}\!+\!\Gamma_{\vec{k}2}\right)\right]^{2} - \gamma_{\vec{k}}^{2}}
	\mp \eta \left(\Gamma_{\vec{k}2}-\Gamma_{\vec{k}1}\right),
	\label{eq:magsu2}
\end{align}
While degenerate for $J_2=0$, these magnon bands clearly split once $J_2 \neq 0$. As illustrated in Fig.~\ref{fig:spinw1}(a,b), the splitting occurs for every momentum except the lines $k_x = 0$ and $k_y=0$, where the degeneracy is enforced by the invariance of $\HH_S$ and $\ket{\psi_0}$ under transformations that combine spin inversion with a mirror operation that exchanges sublattices $A$ and $B$ \cite{smejkal22}.
The index $s$ in $\omega_{\vec{k}s}$ is a quantum number representing the amount by which the total magnetization changes after the creation of a single magnon: $\Delta S_\mathrm{tot}^z = s$ with $s=\pm 1$ \cite{costa24}. We will see below that $s$ is directly related to the chirality of the magnon.

To investigate the excitations of the SU(3) model, we employed linear flavor-wave theory \cite{papa88,joshi99,laeuchli06,tsunetsugu06,bauer12}. Formally, this method is based on a $1/M$ expansion in which fluctuations about the large-$M$ ground state $\ket{\psi_0}$ are described by introducing two flavors of bosons $a_{i\mu\nu}$ per site $(i\mu)$ with $\nu\neq\mu$. The latter obey $a_{i\mu\nu}\ket{\psi_0}=0$ and, along with their creation counterparts $a_{i\mu\nu}^\dagger$, promote transitions between the elements of the local basis formed by the eigenstates of the $S_{i\mu}^{\alpha\alpha}$ operators. By applying $a_{i\mu\nu}^\dagger$ to such a state, one changes its eigenvalues with respect to $S_{i\mu}^{\nu\nu}$ by $+1$ $(S_{i\mu}^{\mu\mu}$ by $-1)$. As derived in the SM \cite{suppl}, the bilinear part of $\HH_S$ reads
\begin{equation}
	\frac{\mathcal{H}_S^{(2)}}{J_1 M} = \sum_{\vec{k} \mu} \sum_{\nu>\mu}
	\Psi_{\vec{k},\mu\nu}^\dagger
	\begin{pmatrix}
		1 - 2\tilde{\eta} \tilde{\Gamma}_{\vec{k}\mu} & \tilde{\gamma}_{\vec{k},\mu\nu} \\
		\tilde{\gamma}^*_{\vec{k},\mu\nu} & 1 - 2\tilde{\eta} \tilde{\Gamma}_{\vec{k}\nu}
	\end{pmatrix}
	\Psi_{\vec{k},\mu\nu}
	\label{eq:H preBog SU3}
\end{equation}
with $\Psi_{\vec{k}\mu}^\dagger = \begin{pmatrix} a_{\vec{k},\mu\nu}^\dagger & a_{-\vec{k},\nu\mu} \end{pmatrix}$ and $\tilde{\eta} = J_{2}/\left(3J_{1}\right)$. Furthermore, $\tilde{\gamma}_{\vec{k},\mu\nu} = \frac{1}{3}\sum_{\boldsymbol{\delta}_{\mu\nu}} e^{i\vec{k} \cdot \boldsymbol{\delta}_{\mu\nu}}$ and $\tilde{\Gamma}_{\vec{k}\mu} = 1 - \cos\left(\vec{k} \cdot \boldsymbol{\Delta}_{\mu} \right)$, where $\boldsymbol{\delta}_{\mu\nu}$ represents the three vectors directed from a site $\left(i\mu\right)$ to its nearest neighbors on sublattice $\nu$ and $\boldsymbol{\Delta}_\mu$ is a vector separating two sites on sublattice $\mu$ that are connected by a $J_2$ bond.
After a Bogoliubov transformation, one obtains $N(N-1) = 6$ magnon bands \cite{suppl}
\begin{align}
	\frac{\omega_{\vec{k},\mu\nu}}{J_1M} &=
	\sqrt{ \left[1 - \tilde{\eta}
		\left(\tilde{\Gamma}_{\vec{k}\mu} + \tilde{\Gamma}_{\vec{k}\nu}\right)
		\right]^{2} - \abs{ \tilde{\gamma}_\vec{k} }^{2} }
	\notag \\
	&+ \sgn(\nu-\mu) \, \tilde{\eta} \left(\tilde{\Gamma}_{\vec{k}\mu} - \tilde{\Gamma}_{\vec{k}\nu}\right)
	\label{eq:magsu3}
\end{align}
with $\mu \ne \nu$. For $J_2=0$, Eq.~\eqref{eq:magsu3} agrees with earlier results \cite{tsunetsugu06,bauer12} and yields a sixfold degeneracy in the reduced Brillouin zone. However, as shown in Fig.~\ref{fig:spinw1}(c,d), the magnon bands once again split for $J_2 \ne 0$. Partial degeneracies, which are protected by spin-space symmetries \cite{suppl}, occur along lines with polar angle $\phi_\vec{k} = n\pi/6$ and at the boundaries of the Brillouin zone.

It is important to note that Eq.~\eqref{eq:H preBog SU3} has three separate $2\times2$ blocks instead of a full $6\times6$ matrix per momentum $\vec{k}$. This simplification follows from the invariance of $\HH_S$ and $\ket{\psi_0}$ (up to a global phase) under SU(3) transformations generated by $S_\mathrm{tot}^{\alpha\alpha} = \sum_j S_j^{\alpha\alpha}$. Indeed, one can verify that the only bosonic bilinears involving $a_{\vec{k}\mu\nu}$ that simultaneously conserve all $S_\mathrm{tot}^{\alpha\alpha}$ and the total momentum are $a_{\vec{k}\mu\nu}^\dagger a_{\vec{k}\mu\nu}$ and $a_{-\vec{k},\nu\mu} a_{\vec{k}\mu\nu}$. This constrained structure implies that, as in the SU(2) case, the creation of a magnon of energy $\omega_{\vec{k},\mu\nu}$ on top of an eigenstate of $\HH^{(2)}_S$ leads to a quantized change $\Delta S_\mathrm{tot}^{\alpha\alpha} = \delta_{\alpha\mu} - \delta_{\alpha\nu}$.
Similar considerations apply for larger $N$ and indicate that the $1/M$ expansion about an $N$-color state results in a linear flavor-wave Hamiltonian with $\binom{N}{2} = N(N-1)/2$ two-dimensional blocks for every $\vec{k}$. Each block yields a pair of magnon modes, with dispersions $\omega_{\vec{k},\mu\nu}$ and $\omega_{\vec{k},\nu\mu}$, that only affect spins on sublattices $\mu$ and $\nu$ and have opposite sets of quantum numbers $\Delta S_\mathrm{tot}^{\alpha\alpha}$ \cite{suppl,footnote1}.

In the SM \cite{suppl}, we show that SU($N>2$) models generally have nonvanishing three-boson terms in their flavor-wave Hamiltonians. This indicates that generalized altermagnets are susceptible to magnon decay processes which are absent from their collinear SU(2) counterparts.


\paragraph{Magnon chirality. --}
Consider an excited state containing magnons of a single momentum and species, but which is otherwise arbitrary. For the SU(2) model, this would be $\ket{\phi_{\vec{k}s}} = \sum_{n=0}^{\infty} c_{n} \ket{n_{\vec{k}s}}$, where $\ket{n_{\vec{k}s}}$ represents a state with $n$ magnons of energy $\omega_{\vec{k}s}$. We define the chirality $\kappa_s$ of the corresponding magnon mode as the sense of precession (as a function of time) of an individual spin around its classical axis, $\unitvec{z}$, in the state $\ket{\phi_{\vec{k}s}}$:
\begin{equation}
\kappa_{s} = \sgn\left\{
\left[ \expval{\vec{S}_{iA}\left(t\right)} \times\frac{d}{dt} \expval{\vec{S}_{iA}\left(t\right)} \right] \cdot \unitvec{z} \right\}.
\label{eq:kappa SU2}
\end{equation}
Here, $\expval{\cdots} = \mel{\phi_{\vec{k}s}}{\cdots}{\phi_{\vec{k}s}}$, $\vec{S}_{iA}\left(t\right) = e^{i\mathcal{H}_S^{\left(2\right)}t} \vec{S}_{iA} e^{-i\mathcal{H}_S^{\left(2\right)}t}$, and the site $(iA)$ is chosen arbitrarily. Using the eigenstates of Eq.~\eqref{eq:H preBog SU2}, a straightforward calculation yields \cite{suppl}
\begin{equation}
\kappa_s = s,
\label{eq:magchir}
\end{equation}
provided that $\xi = \sum_{n=0}^{\infty} \sqrt{n+1} c_n^{*} c_{n+1}\ne0$; otherwise, $\kappa_{s}=0$. The latter case arises, e.g., when $\ket{\phi_{\vec{k}s}}$ has a fixed number of magnons, $c_{n} = \delta_{nm}c_{m}$, and uncertainty relations prevent the determination of the phase of the precessive motion \cite{carruthers68}.
By excluding the possibility $\kappa_{s} = 0$, we thus can assign a definite chirality to each magnon mode. When $J_2 \ne 0$, the energies $\omega_{\vec{k}s}$ of modes with different chiralities $\kappa_{s}$ split, as expected for an SU(2) altermagnet \cite{smejkal23}.
For completeness, we note that the quantity $\tilde{\kappa}_s$ defined by replacing $(iA)$ in Eq.~\eqref{eq:kappa SU2} with a site on the $B$ sublattice would obey $\tilde{\kappa}_s=-s$. This simply reflects the fact that $A$ and $B$ spins precess in phase but with opposite classical ordering directions.

In the SU(3) case, we can similarly consider $\ket{\phi_{\vec{k},\mu\nu}} = \sum_{n=0}^{\infty} c_n \ket{n_{\vec{k},\mu\nu}}$, which is a linear combination of states with $n$ magnons of energy $\omega_{\vec{k},\mu\nu}$. The block-diagonal structure of Eq.~\eqref{eq:H preBog SU3} implies, however, that the expectation values $\expval{S_{i\mu}^{\alpha\beta} (t)}{\phi_{\vec{k},\mu\nu}}$ only depend explicitly on time if $(\alpha,\beta) = (\mu,\nu)$ or $(\nu,\mu)$. Thanks to this property, one can represent the dynamics of $S_{i\mu}^{\alpha\beta}$ in $\ket{\phi_{\vec{k},\mu\nu}}$ and $\ket{\phi_{\vec{k},\nu\mu}}$ in terms of the motion of three-dimensional vectors on the same spherical manifold \cite{suppl}. By using this fact and adopting a definition of analogous to Eq.~\eqref{eq:kappa SU2}, we find that the excitations of our SU(3) model -- and, in fact, of any SU($N$) model constructed through the procedure above \cite{footnote2} -- occur in \emph{pairs} of opposite chiralities. And yet, while bands with different chiralities do split once $J_2 \ne 0$, to define an SU($N$) altermagnet in these terms would be to neglect the fact that its spectrum is \emph{fully} nondegenerate. Hence, we conclude that an insulating SU($N$) altermagnet is (more fundamentally) characterized by the splitting of magnon bands with different sets of quantum numbers $\Delta S_\mathrm{tot}^{\alpha\alpha}$.


\begin{figure}[!tb]
\includegraphics[width=\columnwidth]{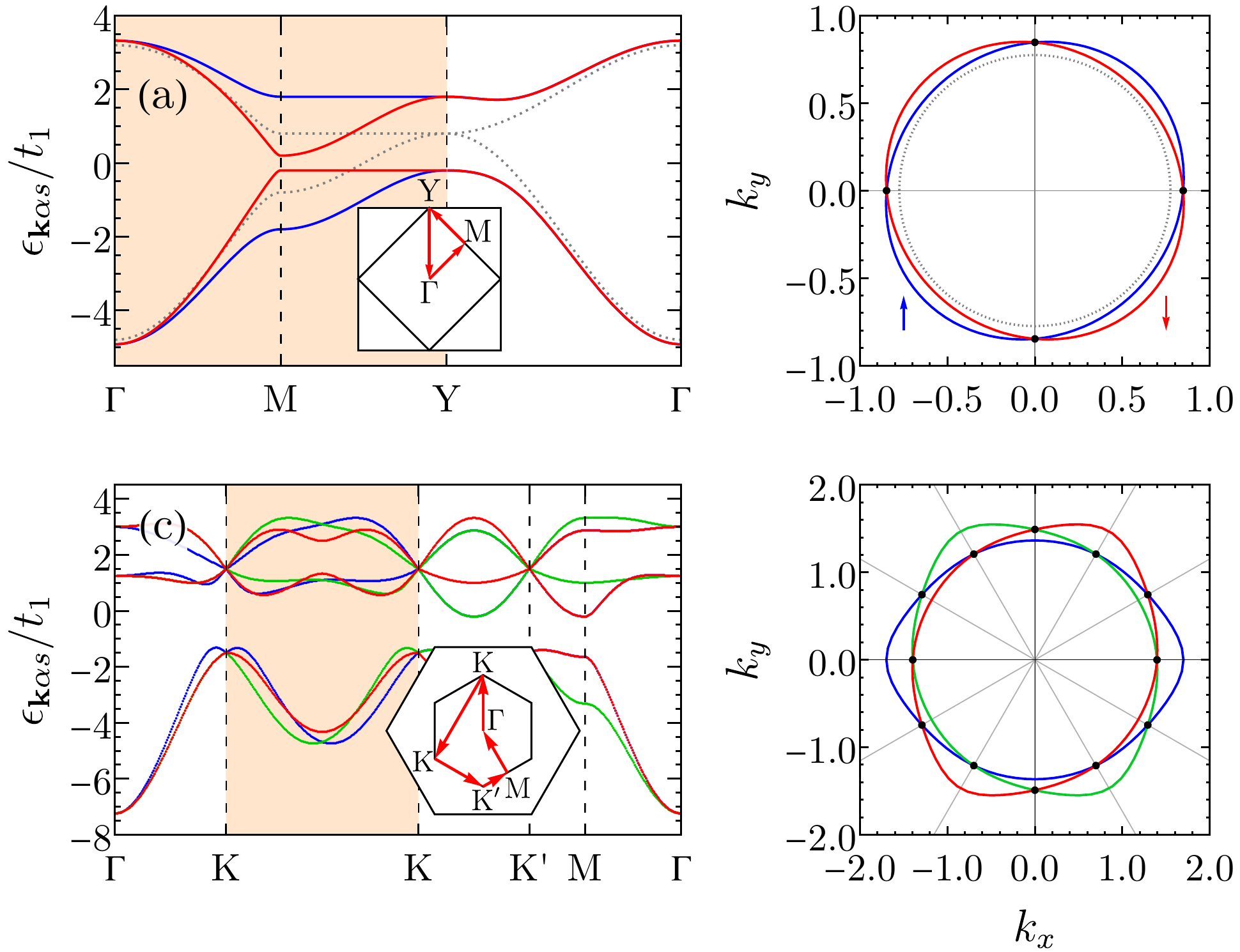}
\caption{Band structure of the (a,b) SU(2) and (c,d) SU(3) double-exchange models for parameters $\left(t_2, K\right) = \left(0.5,5\right)t_1$ and $\left(t_2, K\right) = \left(0.5,6.0\right)t_1$, respectively. The bands are colored according to their spin or flavor. In the left panels, the orange shading highlights sections with completely split bands. The dotted gray lines in (a) and (b) indicate the unsplit bands for $K=0$. The right panels show isoenergy contours for (b) $E = -4.3 t_1$ and (d) $E = -3.7 t_1$.}
\label{fig:bands1}
\end{figure}

\paragraph{Electronic band structure. --}
To extend our considerations to band-electron properties, we supplement the Hamiltonians $\HH_S$ above with a local coupling between the magnetic moments and conduction electrons that hop on the same lattice. We moreover restrict the hopping amplitudes $t_{jj'}$ to first- and second-neighbor sites and require that they obey the same spatial symmetries as $J_{jj'}$. The full Hamiltonian then reads $\mathcal{H} = \mathcal{H}_S + \mathcal{H}_e$ with
\begin{align}
\label{eq:hde}
\mathcal{H}_e
&= - \sum_{jj'} t_{jj'} c_{j \alpha}^\dagger c_{j'\alpha}
- \frac{K}{M} \sum_{j} \sum_{n=1}^{N^2-1} S_{j}^{n} c^\dagger_{j\alpha} \tau^{n}_{\alpha\beta}c_{j\beta}
\end{align}
being the SU($N$) generalization of a standard double-exchange \cite{anderson55,degennes60} or Kondo-lattice \cite{hewson,parcollet98} Hamiltonian. The symbols $\tau^{n}$ ($S_{j}^{n}$) represent the $N^2-1$ independent SU($N$) generators acting on the electronic (spin) sector. Assuming that the conduction electrons transform under the fundamental representation of SU($N$), we can identify $\tau^{n}$ as one-half of the Pauli (Gell-Mann) matrices for $N=2$ ($3$). The $S_{j}^{n}$, in contrast, are assumed to be in the $\left\{M,0\right\}$ irrep as before. $(N-1)$ of these operators are diagonal and given by linear combinations of the $S_j^{\alpha\alpha}$, while the remaining $N(N-1)$ can be combined to form ladder operators $S_j^{\alpha\beta}$ with $\alpha\ne\beta$ \cite{suppl}.

We now assume that the local moments $S_{j}^{n}$ display long-range altermagnetic order with an $N$-site unit cell and treat the model \eqref{eq:hde} within a semiclassical approximation that replaces each $S_j^{n} \to h_{j}^n = \expval{S_{j}^{n}}{\psi_0}$. Thus, the electrons experience an effective field $h_{j}^n$ and form $N^2$ bands in the reduced Brillouin zone. Since $\HH$ and $\ket{\psi_0}$ are invariant under global transformations generated by the $(N-1)$ diagonal SU($N$) generators, the mean-field Hamiltonian $\HH_e^\mathrm{MF}$ splits into $N$ separate $N \times N$ blocks, each corresponding to a different SU($N$) flavor $\alpha$.

In the SU(2) checkerboard model, $h_j^{n}$ takes the form of a staggered N\'eel field, and the diagonalization of the hopping problem yields the band structure
\begin{align}
\epsilon_{\vec{k}\alpha\pm} &= -2t_{2} \cos k_{x} \cos k_{y}
\nonumber \\
&\pm \sqrt{\left[2t_{2}\sin k_{x}\sin k_{y} + \left(-1\right)^{\alpha} K/4 \right]^{2} + \left(4t_{1}\gamma_{\vec{k}}\right) ^2 }.
\end{align}
Thus, each pair of bands $\epsilon_{\vec{k}\alpha s}$ with opposite spin $\alpha$ splits when $t_2,K \ne 0$ [see Fig.~\ref{fig:bands1}(a,b)]. This splitting further displays a $d$-wave symmetry, as required by spin-space symmetries and expected for $d$-wave altermagnets \cite{smejkal22}.

For the SU(3) model, we determine the dispersion by diagonalizing each $3\times3$ block in $\mathcal{H}_e$ numerically. Similarly to the SU(2) case, we find three different sets of flavor-degenerate bands when either $t_2=0$ or $K=0$, since the system has a three-site unit cell in both situations. However, as illustrated in Fig.~\ref{fig:bands1}(c,d), $t_2, K \ne 0$ induces a complete splitting of the nine bands. Partial degeneracies, protected by spin point-group symmetries, are retained along high-symmetry lines as before \cite{suppl}.


\paragraph{Conclusion and outlook. --}
In summary, we have shown that SU($N\!>\!2$) magnets can display a generalized form of altermagnetism. We made this point by constructing SU($N$)-symmetric models which, while having ground states with $N$-color order, exhibit hallmarks of SU(2) altermagnetism in their excitation spectra. Specifically, insulating versions of the models feature nondegenerate magnon bands with different sets of quantum numbers, whereas their metallic counterparts have flavor-split electronic bands. Our construction underlines that magnon quantum numbers are more fundamental to the splitting of altermagnetic magnon bands than the concept of magnon chirality.

Our work suggests several avenues for future theoretical investigation. Models with large $N$ may provide a useful starting point to the study quantum phase transitions via a controlled $1/N$ expansion. Given that SU(2) altermagnets display large anomalous Hall \cite{nagaosa10,smejkal20} and Nernst effects \cite{smejkal_rev}, it would also be interesting to search for unconventional transport properties in SU($N$) generalizations. And finally, instabilities out of SU($N$) altermagnets may lead to new exotic phases of matter.

While systems with orbital degeneracy can realize higher SU($N\!>\!2$) symmetries \cite{kugel73,joshi99}, reproducing this property in a solid whose magnetic interactions and spatial symmetries are compatible with altermagnetism is a daring task. An alternative, and perhaps more realistic, route to an experimental realization of our ideas is to engineer the models presented here in systems of ultracold alkaline-earth atoms, for which nuclear spins can display SU($N$)-symmetric interactions with $N\le10$ \cite{gorshkov10,pagano14,zhang14,hofrichter16,rey_rop14,taie24}. Similarly to a recent proposal for the simulation of an SU(2) Hubbard model on the checkerboard lattice \cite{knap23}, which reduces to a Heisenberg Hamiltonian \eqref{eq:heis} in the strong-coupling limit, we propose to implement an SU(3) hexatriangular Hubbard model by superimposing different triangular optical lattices \cite{schauss23}.


\acknowledgments

We thank S.-W. Cheong for helpful discussions.
Financial support from the DFG through SFB 1143 (project-id 247310070) and the W\"urzburg-Dresden Cluster of Excellence on Complexity and Topology in Quantum Matter -- \textit{ct.qmat} (EXC 2147, project-id 390858490) is gratefully acknowledged.



\begin{thebibliography}{}
\bibitem{footnote1}
Under a suitable change of basis, one can in fact each magnon band with a different root of the SU($N$) algebra.
\bibitem{footnote2}
This is related to the fact that one can form $\binom{N}{2}$ SU(2) subalgebras of SU($N$) by combining pairs of ladder operators corresponding to opposite roots, $S^{\mu\nu}$ and $S^{\nu\mu}$, with a suitable linear combination of elements of the Cartan subalgebra \cite{zee_book}.
\end{thebibliography}
\end{document}


\title{Supplemental material for: \\
SU($N$) altermagnetism: Lattice models, magnon modes, and flavor-split bands}
\author{Pedro M. C\^onsoli}
\author{Matthias Vojta}
\affiliation{Institut f\"ur Theoretische Physik and W\"urzburg-Dresden Cluster of Excellence ct.qmat, Technische Universit\"at Dresden, 01062 Dresden, Germany}

\date{\today}
\maketitle


\section{SU(4) model on a ``cross-cubic'' lattice}

To substantiate the claim that the construction principle described in the main text can be applied beyond $N=2$ and $3$, we discuss in this section possible implementations of an SU(4) lattice model with an altermagnetic phase.

\begin{figure}
	\centering
	\includegraphics[width=0.75\columnwidth]{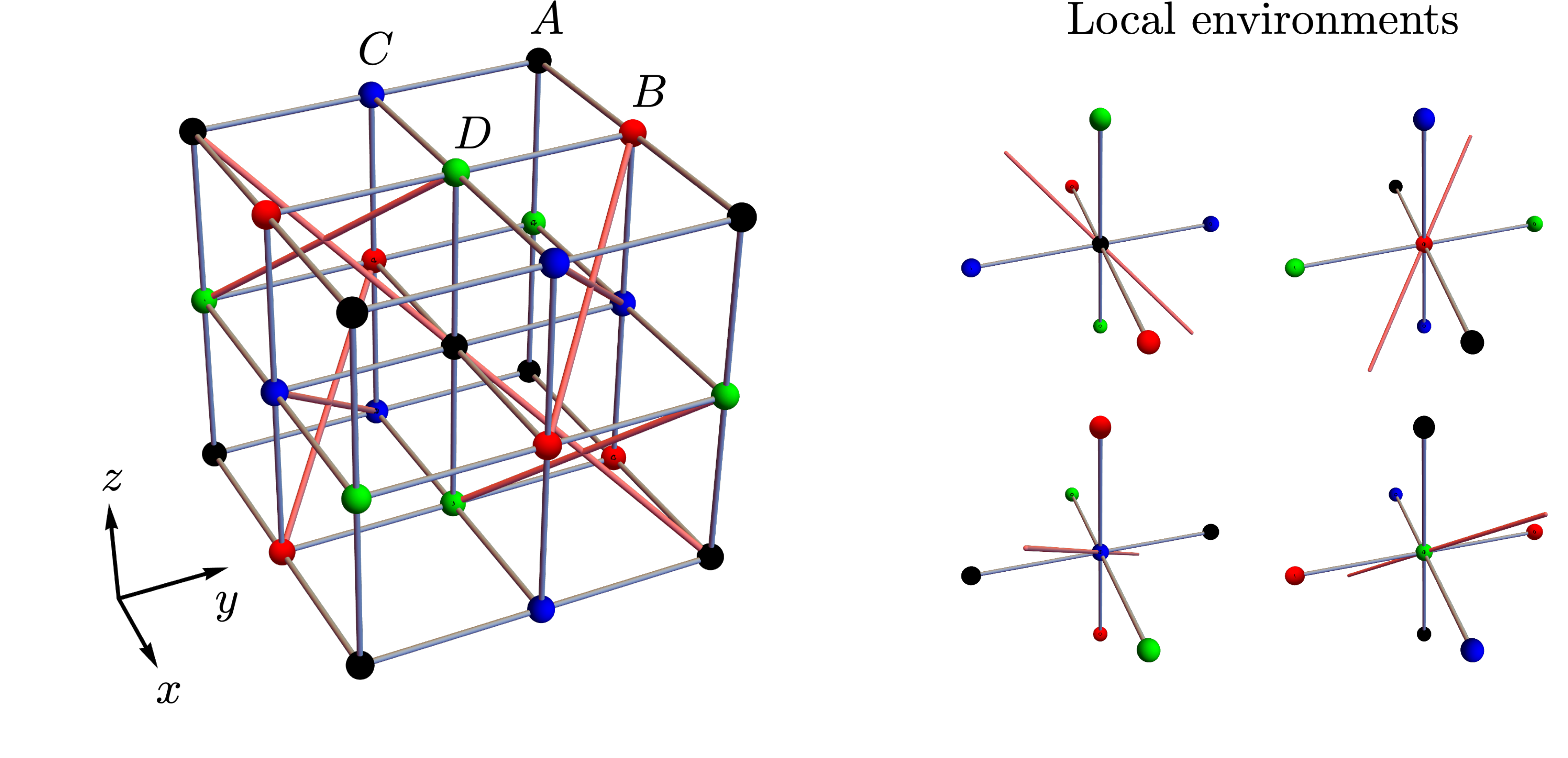}
	\caption{Section of a ``cross-cubic'' lattice with four-color order which realizes an SU(4) altermagnet. The displacement vectors $\bm{\Delta}_\mu$ given in text refer to the sublattices as labeled on the left. On the right, we show the four different types of local environments.}
	\label{fig:crosscubic}
\end{figure}

For step (i), we consider a simple cubic lattice, which admits a quadripartite decomposition into four body-centered cubic sublattices $\left\{1,2,3,4\right\} \equiv \left\{A,B,C,D\right\}$, as illustrated in Fig.~\ref{fig:crosscubic}. We then posit that there exists an SU(4)-symmetric model that respects all spatial symmetries of the cubic lattice and whose ground state is adiabatically connected to the four-color product state, $\ket{\psi_0}$, given by Eq.~(2) in the main text with $N=4$.
This is certainly fulfilled by a SU(4) Heisenberg model, see Eq.~(1) in the main text, with a nearest-neighbor antiferromagnetic coupling $J_1$ and sufficiently large $M$. However, to the best of our knowledge, there is currently no evidence that it holds for the fundamental representation, $M=1$.
Alternatively, one could start from a related SU(4)-symmetric model introduced in Sec.~6.B. of Ref.~\cite{kiselev16}. Using Monte Carlo simulations, the authors proved that the system displays a four-color ordered ground state for every value of $M$ compatible with their model instead of realizing a simplex solid \cite{arovas08}.

For step (ii), we lower the symmetry of the cubic lattice by adding selected third-neighbor couplings, $J_3$, depicted as pink lines in Fig.~\ref{fig:crosscubic}. In the coordinate system shown in that same figure, the $J_3$ couplings occur along the directions $\bm{\Delta}_A = \left(-1,-1,1\right)$, $\bm{\Delta}_B = \left(-1,1,1\right)$, $\bm{\Delta}_C = \left(1,-1,1\right)$, and $\bm{\Delta}_D = \left(1,1,1\right)$. This eliminates all inversion and translation symmetries connecting the different sublattices, but preserves at least a $C_4$ symmetry with respect to an axis that is parallel to $\unitvec{z}$ and goes through the centers of empty cubes stacked vertically. Due to this symmetry breaking, the system will display split magnon bands without an net SU(4) moment, and thus consist of an altermagnet.

Based on this and models considered in the main text, the generalization of the construction principle to larger $N$ seems to be straightforward if a given lattice $\mathcal{L}$ admits an $N$-partite decomposition and each sublattice $\mu$ is itself a lattice with coordination number $2N$.

\section{SU(2) local-moment altermagnet on the checkerboard lattice}
\label{sec:SU2ins}

In this section, we present details on the spin-wave calculations we performed for the SU(2) Heisenberg model considered in the main text. This includes a discussion of the symmetries of the magnon spectrum and a derivation of the chirality of the magnon modes. The Hamiltonian
\begin{equation}
\HH_S =
J_1\sum_{\left\langle i\mu,j\nu\right\rangle }\vec{S}_{i\mu}\cdot\vec{S}_{j\nu}
+ J_2\sum_{\langle\!\langle i\mu,j\mu\rangle\!\rangle}\vec{S}_{i\mu}\cdot\vec{S}_{j\mu}
\label{eq:H checkerboard}
\end{equation}
has spins $S$ distributed on the sites $(i\mu)$ of a checkerboard lattice with $N_{s} = 2N_{c}$ sites and magnetic couplings $J_1>0$ ($J_2$) represented by the solid (dashed) lines in Fig.~1(a) of the main paper.


\subsection{Spin-wave theory}
\label{subsec:su2swt}

Assuming collinear Néel order and introducing Holstein-Primakoff (HP) bosons $a_i$ and $b_i$ for sublattices $A$ and $B$, respectively, we expand the Hamiltonian $\HH_S$ in powers of $1/\sqrt{S}$. By truncating at order $S$ (quadratic in bosons) and applying Fourier transforms with the convention $a_{i} = N_c^{-1/2} \sum_{\vec{k}} e^{i\vec{k}\cdot\vec{r}_{iA}} a_{\vec{k}}$, we arrive at the linear spin-wave Hamiltonian
\begin{equation}
	\HH_{\mathrm{LSW}} = N_{s} S\left(S+1\right)\left(J_2-2J_1\right) + \sum_\vec{k}
	\begin{pmatrix}
		a_\vec{k}^\dagger & b_{-\vec{k}}
	\end{pmatrix} 
	\begin{pmatrix}
		A_\vec{k} & C_\vec{k} \\
		C_\vec{k}^* & B_\vec{k}
	\end{pmatrix}
	\begin{pmatrix}
		a_\vec{k} \\
		b_{-\vec{k}}^\dagger
	\end{pmatrix},
	\label{eq:h02}
\end{equation}
where
\begin{align}
	A_\vec{k} &= 4J_1 S (1 - 2\eta \Gamma_{\vec{k}1}), &
	B_\vec{k} &= 4J_1 S (1 - 2\eta \Gamma_{\vec{k}2}), &
	C_\vec{k} &= 4J_1 S \gamma_\vec{k}.
\end{align}
The quantities $\eta$, $\gamma_\vec{k}$, and $\Gamma_{\vec{k}\mu}$ are defined as in the main text, so that the bilinear part of $\HH_{\mathrm{LSW}}$ is identical to Eq.~(3). After diagonalization via a Bogoliubov transformation, Eq.~\eqref{eq:h02} becomes
\begin{equation}
	\HH_{\mathrm{LSW}} = N_s S\left(S+1\right)\left(J_2-2J_1\right)
	+ \sum_{\vec{k}} \sum_{s=\pm} \omega_{\vec{k}s}\left(\alpha_{\vec{k}s}^{\dagger}\alpha_{\vec{k}s}+\frac{1}{2}\right).
	\label{eq:H postBog SU2}
\end{equation}
The dispersion
\begin{equation}
	\omega_{\vec{k} s} = \sqrt{ \left( \frac{A_\vec{k} + B_\vec{k}}{2} \right)^2 - \abs{C_\vec{k}}^2 } - s \left( \frac{A_\vec{k} - B_\vec{k}}{2} \right)
	\label{eq:disp SU2}
\end{equation}
yields Eq.~(4) of the main text, and the magnon eigenmodes
\begin{equation}
	\begin{pmatrix}
		\alpha_{\vec{k},-} \\
		\alpha_{-\vec{k},+}^{\dagger}
	\end{pmatrix}
	=
	\begin{pmatrix}
		u_{\vec{k}} & -v_{\vec{k}} e^{i\varphi_\vec{k}} \\
		-v_{\vec{k}} e^{-i\varphi_\vec{k}} & u_{\vec{k}}
	\end{pmatrix}
	\begin{pmatrix}
		a_{\vec{k}} \\
		b_{-\vec{k}}^{\dagger}
	\end{pmatrix},
\label{eq:Bog SU2}
\end{equation}
are given in terms of $\varphi_\vec{k} = \arg C_\vec{k}$ (which vanishes in this case, since $C_\vec{k}$ is real) and the Bogoliubov coefficients
\begin{align}
	u_{\vec{k}} &= 
	\cosh \left[ - \frac{1}{2} \tanh^{-1} \left( \frac{2 \abs{C_\vec{k}}}{A_\vec{k} + B_\vec{k}} \right) \right]
	=
	\sqrt{ \frac{4J_1-J_2\left(\Gamma_{\vec{k}1}+\Gamma_{\vec{k}2}\right)}{\left(\omega_{\vec{k}+}+\omega_{\vec{k}-}\right)/S} + \frac{1}{2}},
	\label{eq:uk SU2}
	\\
	v_{\vec{k}} &= 
	\sinh \left[ - \frac{1}{2} \tanh^{-1} \left( \frac{2 \abs{C_\vec{k}}}{A_\vec{k} + B_\vec{k}} \right) \right]
	=-\sqrt{ \frac{4J_1-J_2\left(\Gamma_{\vec{k}1}+\Gamma_{\vec{k}2}\right)}{\left(\omega_{\vec{k}+}+\omega_{\vec{k}-}\right)/S} - \frac{1}{2}}.
	\label{eq:vk SU2}
\end{align}
Here, one can see that both $u_\vec{k}$ and $v_\vec{k}$ diverge as $\left(\omega_{\vec{k}+}+\omega_{\vec{k}-}\right)^{-1/2}\sim\omega_{\vec{k}s}^{-1/2}$ at the Goldstone wavevector $\vec{k}=\vec{0}$. This implies that the products $u_{\vec{k}}^{2}\omega_{\vec{k}s}$ and $v_{\vec{k}}^{2}\omega_{\vec{k}s}$ are always finite and strictly positive, a property we will invoke in the derivation of the magnon chirality in Sec.~\ref{sec:su2chirality}.

Finally, we note that the structure of Eq.~\eqref{eq:Bog SU2} is such that the creation of a single $\alpha_{\vec{k}s}^\dagger$ changes the total magnetization of any eigenstate of $\HH_{\mathrm{LSW}}$ by a quantized amount $\Delta S^z_\mathrm{tot} = s$. As mentioned in the main text, this is a consequence of the invariance of $\HH_S$ and the Néel product state, $\ket{\psi_0}$, under global rotations generated by $S^z_\mathrm{tot}$.


\subsection{Symmetries of the spin-wave spectrum}
\label{subsec:symmetries lsw}

In this subsection, we show why eliminating the inversion and translation symmetries between the different magnetic sublattices leads to the splitting of the magnon bands. Our analysis relies of the identification of symmetries of $\HH_{\mathrm{LSW}}$ which, in addition to commuting with the Hamiltonian \eqref{eq:H checkerboard}, leave the reference state $\ket{\psi_0}$ for the spin-wave expansion (a Néel state order along the $z$ axis) invariant.

Let $\ssop{A}{B}$ denote a transformation that acts on spin space with $A$ and in real space with $B$. Since we ignore relativistic effects, $A$ and $B$ are completely independent, except for when time reversal is applied. In that case, the presence of the time-reversal operator $\Theta$ in $B$ must be accompanied by spin inversion $\bar{E}$ in $A$.
Due to the collinearity of the Néel state, one of the symmetries of $\HH_{\mathrm{LSW}}$ is $T_0 = \ssop{ \bar{E} C_{2} }{\Theta}$, where $C_2 = e^{-i\pi S_x}$ implements a 180° rotation around the $x$-axis in spin space.
There are, in addition, unitary symmetries $T = \ssop{C_2}{ \left\{ R | \vec{t} \right\} }$ that swap the magnetic sublattices via a combination of a (proper or improper) rotation $R$ and a translation $\vec{t}$. The action of these transformations on the spin operators is such that
\begin{align}
	T_0 \left( S_{iA}^z, S_{iA}^{\pm} \right) T_0^{-1} &= \left( S_{iA}^z, -S_{iA}^{\pm} \right) ,
	&
	T \left( S_{iA}^z, S_{iA}^{\pm} \right) T^{-1} &= \left( -S_{jB}^z, S_{jB}^{\mp} \right),
\end{align}
with $\vec{r}_{jB} = R[\vec{r}_{iA}] + \vec{t}$. By employing the HP transformation, we then find that
\begin{align}
	T_0 a_\vec{k} T_0^{-1} &= -a_{-\vec{k}},
	&
	T a_\vec{k} T^{-1} &= e^{i R[\vec{k}] \cdot \vec{t}} b_{R[\vec{k}]}.
\end{align}

With previous results, one can compute the action of $T_0$ on the bilinear part of $\HH$ as follows:
\begin{align}
	T_0 \HH_S^{(2)} T_0^{-1} &= \sum_{\vec{k}}
	\begin{pmatrix}
		a_{-\vec{k}}^\dagger & b_{\vec{k}}
	\end{pmatrix}
	\begin{pmatrix}
		A_\vec{k} & C_\vec{k}^* \\
		C_\vec{k}	& B_\vec{k}
	\end{pmatrix}
	\begin{pmatrix}
		a_{-\vec{k}} \\
		b_{\vec{k}}^\dagger
	\end{pmatrix}
	\notag \\
	&= \sum_{\vec{k}}
	\begin{pmatrix}
		a_{\vec{k}}^\dagger & b_{-\vec{k}}
	\end{pmatrix}
	\begin{pmatrix}
		A_{-\vec{k}} & C_{-\vec{k}}^* \\
		C_{-\vec{k}}	& B_{-\vec{k}}
	\end{pmatrix}
	\begin{pmatrix}
		a_{\vec{k}} \\
		b_{-\vec{k}}^\dagger
	\end{pmatrix}
	=
	\sum_{\vec{k}s} \omega_{-\vec{k},s} \alpha_{\vec{k}s}^\dagger \alpha_{\vec{k}s},
\end{align} 
where we used Eq.~\eqref{eq:disp SU2} in the last equality. Since $T_0 \HH T_0^{-1} = \HH$, we conclude that $\omega_{\vec{k}s} = \omega_{-\vec{k},s}$, i.e., $T_0$ symmetry implies that the dispersion is inversion-symmetric. We can similarly analyze the consequences of a unitary symmetry $T$ by evaluating
\begin{align}
	T \HH_S^{(2)} T^{-1} &= \sum_{\vec{k}}
	\begin{pmatrix}
		e^{ -i R[\vec{k}] \cdot \vec{t} } b_{R[\vec{k}]}^\dagger & e^{ i R[\vec{k}] \cdot \vec{t} } a_{-R[\vec{k}]}
	\end{pmatrix}
	\begin{pmatrix}
		A_\vec{k} & C_\vec{k} \\
		C_\vec{k}^*	& B_\vec{k}
	\end{pmatrix}
	\begin{pmatrix}
		e^{ i R[\vec{k}] \cdot \vec{t} } b_{R[\vec{k}]} \\
		e^{ -i R[\vec{k}] \cdot \vec{t} } a_{-R[\vec{k}]}^\dagger
	\end{pmatrix}
	\notag \\
	&= \sum_{\vec{k}}
	\begin{pmatrix}
		a_{-R[\vec{k}]}^\dagger & b_{R[\vec{k}]}
	\end{pmatrix}
	\begin{pmatrix}
		B_\vec{k} & e^{ 2i R[\vec{k}] \cdot \vec{t} } C_\vec{k} \\
		e^{ -2i R[\vec{k}] \cdot \vec{t} } C_\vec{k}^*	& A_\vec{k}
	\end{pmatrix}
	\begin{pmatrix}
		a_{-R[\vec{k}]} \\
		b_{R[\vec{k}]}^\dagger
	\end{pmatrix}
	+ \sum_\vec{k} \left( B_\vec{k} - A_\vec{k} \right)
	\notag \\
	&= \sum_\vec{k}
	\begin{pmatrix}
		a_{\vec{k}}^\dagger & b_{-\vec{k}}
	\end{pmatrix}
	\begin{pmatrix}
		B_{-R^{-1}[\vec{k}]} & e^{ 2i \vec{k} \cdot \vec{t} } C_{-R^{-1}[\vec{k}]} \\
		e^{ - 2i \vec{k} \cdot \vec{t} } C_{-R^{-1}[\vec{k}]}^*	& A_{-R^{-1}[\vec{k}]}
	\end{pmatrix}
	\begin{pmatrix}
		a_{\vec{k}} \\
		b_{-\vec{k}}^\dagger
	\end{pmatrix}
	\notag \\
	&= \sum_{\vec{k}s} \omega_{-R^{-1}[\vec{k}], -s} \alpha_{\vec{k}s}^\dagger \alpha_{\vec{k}s}.
\end{align} 
In going from the first to the second line, we simply rearranged the order of the bosonic bilinears by using their commutation relations. This is the origin of the constant $\sum_\vec{k} \left(B_\vec{k} - A_\vec{k} \right)$, which nonetheless vanishes on account of the rotational symmetry between sublattices $A$ and $B$. The third line is then reached by shifting the momenta in the sum, and the final result follows once again from Eq.~\eqref{eq:disp SU2}. Thus, if $T$ is a symmetry of $\HH$, then
\begin{equation}
	\omega_{R[\vec{k}], s} = \omega_{-\vec{k}, -s}.
	\label{eq:Tsymm SU2}
\end{equation}
In the absence of $J_2$ couplings, both $T_{1} = \left[C_{2} || \left\{\mathcal{P} | \vec{0}\right\}\right]$ and $T_{2} = \left[C_{2} || \left\{E | \bm{\tau} \right\}\right]$, defined in terms of an inversion $\mathcal{P}$ about the center of a nearest-neighbor bond and a translation $\bm{\tau}$ connecting different sublattices, are symmetries of $\HH$ and $\ket{\psi_0}$. According to Eq.~\eqref{eq:Tsymm SU2}, they imply that $\omega_{\vec{k}s} = \omega_{\vec{k},-s}$ and $\omega_{\vec{k}s} = \omega_{-\vec{k},-s}$, respectively. This shows that the degeneracy of magnon bands is enforced by $T_{1}$ alone or by the combination of $T_{2}$ and $T_{0}$.  In the altermagnet, none of these options is realized, leading to the splitting of the magnon bands as discussed in the main text.


\subsection{Magnon chirality}
\label{sec:su2chirality}

The fact that magnons created by $\alpha_{\vec{k}+}^\dagger$ and $\alpha_{\vec{k}-}^\dagger$ have different magnetic quantum numbers is also reflected in the dynamical behavior they generate. To see this, consider a state
\begin{equation}
\Ket{\phi_{\vec{k}s}} = \sum_{n=0}^{\infty} c_{n}\Ket{n_{\vec{k}s}}
= \sum_{n=0}^{\infty}\frac{c_{n}}{\sqrt{n!}}\left(\alpha_{\vec{k}s}^{\dagger}\right)^{n}\Ket{0},
\end{equation}
which is a superposition with arbitrary complex coefficients, $c_n$, of $n$-particle states of a magnon with fixed momentum $\vec{k}$ and quantum number $s$. These are obtained by applying $\alpha_{\vec{k}s}^\dagger$ to the ground state $\Ket{0}$ of the linear spin-wave Hamiltonian, which satisfies $\alpha_{\vec{k}s} \Ket{0} = 0$. Working in the Heisenberg picture, we now examine the time evolution of
\begin{equation}
	\Braket{\vec{S}_{iA}\left(t\right)} = \Re \left[\left\langle S_{iA}^{+} \left(t\right)\right\rangle \left(\unitvec{x} - i\unitvec{y}\right)\right] +\left\langle S_{iA}^{z}\left(t\right)\right\rangle \unitvec{z},
	\label{eq:expval SiA SU2}
\end{equation}
where $\expval{\cdots} = \expval{\cdots}{\phi_{\vec{k}s}}$. By using the Bogoliubov transformation in Eq.~\eqref{eq:Bog SU2} and recalling that $v_{\vec{k}}$ and $\omega_{\vec{k}s}$ are even in $\vec{k}$, we can evaluate the first expectation value in Eq. \eqref{eq:expval SiA SU2} as follows:
\begin{align}
	\expval{ S_{iA}^{+} (t) } &\approx
	\sqrt{\frac{2S}{N_c}} \sum_{m,n} c_{m}^{*}c_{n} \sum_\vec{q} e^{i \vec{q} \cdot \vec{r}_{iA} } \Braket{m_{\vec{k}s} | u_\vec{q} \alpha_{\vec{q}-}(t) + v_\vec{q} \alpha_{-\vec{q},+}^\dagger (t) | n_{\vec{k}s}}
	\nonumber \\
	&= \sqrt{\frac{2S}{N_c}} e^{-si \left(\vec{k}\cdot\vec{r}_{iA}-\omega_{\vec{k}s}t\right) }
	\left[ \delta_{s-} u_\vec{k} \left( \sum_{m=0}^{\infty}\sqrt{m+1} \, c_m^* c_{m+1} \right) + \delta_{s+} v_\vec{k} \left( \sum_{n=0}^\infty \sqrt{n+1} \, c_{n+1}^* c_n \right)\right]
	\nonumber \\
	& =\sqrt{\frac{2S}{N_c}} \left|\xi\right| e^{-si \left(\vec{k} \cdot \vec{r}_{iA} - \omega_{\vec{k}s}t + \arg\xi\right)} \left(\delta_{s-} u_\vec{k} + \delta_{s+} v_\vec{k} \right)
	\label{eq:expval S+ SU2}
\end{align}
with $\xi\left(\left\{ c_{n}\right\} \right)=\sum_{n=0}^{\infty}\sqrt{n+1}\,c_{n}^{*}c_{n+1}$. Similarly,
\begin{align}
	\expval{ S_{iA}^{z} (t) } 
	&= S - \frac{1}{N_{c}} \sum_{\vec{p}\vec{q}} e^{i\left(\vec{p}-\vec{q}\right) \cdot \vec{r}_{iA}} \expval{ a_\vec{q}^{\dagger}(t) a_\vec{p} (t) }
	\nonumber \\
	&= S-\frac{1}{N_{c}} \sum_{m,n} c_{m}^{*}c_{n} \sum_{\vec{p}\vec{q}} \delta_{\vec{pq}}\delta_{mn} \left[\delta_{s-}\delta_{\vec{pk}}u_{\vec{k}}^{2}n + v_{\vec{p}}^{2}\left(1+\delta_{s+}\delta_{\vec{p},-\vec{k}}n\right)\right]
	\nonumber \\
	& =m - \frac{\zeta}{N_{c}} \left(\delta_{s-}u_{\vec{k}}^{2} + \delta_{s+}v_{\vec{k}}^{2}\right),
	\label{eq:expval Sz SU2}
\end{align}
where we have identified the staggered magnetization in the ground state,
\begin{equation}
	m = \frac{1}{N_{c}} \sum_{i\mu}\left(-1\right)^\mu \expval{S_{i\mu}^{z}}{0}
	= S - \frac{1}{N_{c}} \sum_{\vec{k}}v_{\vec{k}}^{2},
	\label{eq:m SU2}
\end{equation}
and defined $\zeta\left(\left\{ c_{n}\right\} \right)=\sum_{n=0}^{\infty}\left|c_{n}\right|^{2}n$. When $\Ket{\phi_{\vec{k}s}}=\Ket{0}$, the latter clearly vanishes and we recover $\expval{S_{iA}^{z} (t)} = m$.

After substituting Eqs.~\eqref{eq:expval S+ SU2} and \eqref{eq:expval Sz SU2} back into Eq. \eqref{eq:expval SiA SU2}, we obtain
\begin{align}
	\expval{ \vec{S}_{iA} (t) } & =\sqrt{\frac{2S}{N_{c}}}
	\abs{\xi} \Re \, \bm{\sigma}_{\vec{k}s}(t)
	+ \left[m-\frac{\zeta}{N_{c}}\left(\delta_{s-}u_{\vec{k}}^{2} + \delta_{s+}v_{\vec{k}}^{2}\right)\right] \unitvec{z} ,
	\label{eq:expval SiA SU2 2}
\end{align}
where
\begin{equation}
	\bm{\sigma}_{\vec{k}s}(t) = e^{-si \left(\vec{k}\cdot\vec{r}_{iA} - \omega_{\vec{k}s}t + \arg\xi\right)} \left(\delta_{s-}u_{\vec{k}} + \delta_{s+}v_{\vec{k}}\right) \left(\unitvec{x}-i\unitvec{y}\right).
\end{equation}
Note that only the tranverse (i.e., $x$ and $y$) components of the spin operator have time-dependent expectation values. This describes the precession of $\expval{ \vec{S}_{iA}\left(t\right) }$ around the ordering axis $\unitvec{z}$, as one can explicitly see by writing
\begin{align}
	\Re \, \boldsymbol{\sigma}_{\vec{k}s} 
	&= \cos\left(\omega_{\vec{k}s}t-\vec{k}\cdot\vec{r}_{iA}-\arg\xi\right)\unitvec{x} 
	-s \sin\left(\omega_{\vec{k}s}t-\vec{k}\cdot\vec{r}_{iA}-\arg\xi\right)\unitvec{y}.
\end{align}
Hence, the sense of the precession depends on $s$, and we conclude that the two magnon modes have different \emph{chiralities}. To sharply draw this distinction, we introduce a ``chirality index''
\begin{align}
	\kappa_{s} & \equiv\sgn\left\{ \left[\expval{ \vec{S}_{iA}\left(t\right) }\times\frac{d}{dt}\expval{ \vec{S}_{iA}\left(t\right) }\right]\cdot\unitvec{z}\right\},
	\label{eq:su2kappa}
\end{align}
which is based on the time evolution of a spin in a semiclassical picture. Since the cross product in Eq. \eqref{eq:su2kappa} is projected onto $\unitvec{z}$, we can simply replace $\vec{S}_{iA}$ by $\vec{S}_{iA}^{\perp} = S_{iA}^x \unitvec{x} + S_{iA}^y \unitvec{y}$ when computing $\kappa_s$. We find that
\begin{align}
	\expval{ \vec{S}_{iA}^{\perp} (t) } \times \frac{d}{dt} \expval{ \vec{S}_{iA}^{\perp} (t) } 
	&= \frac{2S}{N_{c}} \left|\xi\right|^{2} 
	\left(\bm{\sigma}_{\vec{k}s} + \bm{\sigma}_{\vec{k}s}^{*}\right) 
	\times \frac{d}{dt} \left( \bm{\sigma}_{\vec{k}s} + \bm{\sigma}_{\vec{k}s}^{*} \right)
	= -\frac{4S}{N_{c}} \left|\xi\right|^{2} s i \omega_{\vec{k}s} 
	\left( \bm{\sigma}_{\vec{k}s} \times \bm{\sigma}_{\vec{k}s}^{*}\right)
	\nonumber \\
	&= \left[\frac{8S}{N_{c}}\left|\xi\right|^{2}\omega_{\vec{k}s}\left(\delta_{s-}u_{\vec{k}}^{2} + \delta_{s+}v_{\vec{k}}^{2}\right)\right]s\unitvec{z}.
\end{align}
Given that $u_{\vec{k}}^{2}\omega_{\vec{k}s}$ and $v_{\vec{k}}^{2}\omega_{\vec{k}s}$ are strictly positive [see Eqs.~\eqref{eq:uk SU2} and \eqref{eq:vk SU2}], we arrive at
\begin{equation}
	\kappa_{s} = +s \qquad \text{(if \ensuremath{\xi\ne0})}.
	\label{eq:su2kappafinal}
\end{equation}
The condition $\xi\ne0$ is guaranteed to hold, e.g., for a coherent state $\Ket{\phi_{\vec{k}s}} = e^{-\left|\phi\right|^{2}/2}e^{\phi\alpha_{\vec{k}s}^{\dagger}}\Ket{0}$, for which $c_{n} = e^{-\left|\phi\right|^{2}/2}\phi^{n}/\sqrt{n!}$. On the other hand, states with $\xi=0$ have a chirality index $\kappa_{s}=0$. This occurs, in particular, when $\Ket{\phi_{\vec{k}s}}$ has a definite number of magnons, i.e., $c_{n}=\delta_{nm}c_{m}$.
In such a case, the precise knowledge $S_\mathrm{tot}^z$ leads to a maximum uncertainty in $S_\mathrm{tot}^x$ and $S_\mathrm{tot}^y$, which therefore makes the sense of precession generated by the collective modes ill-defined. Finally, note that $\kappa_{s}$ otherwise does not depend on the set of coefficients $\left\{ c_{n}\right\} $ nor on the momentum $\vec{k}$ of the excitation used to construct $\Ket{\phi_{\vec{k}s}}$.


\section{SU(3) local-moment altermagnet on the hexatriangular lattice}
\label{sec:SU3ins}

In this section, we retrace the steps of Sec.~\ref{sec:SU2ins}, but now for the SU(3) Heisenberg model discussed in the main text. We discuss how this SU(3) ``spin'' Hamiltonian emerges as the effective low-energy description in the strong-coupling limit of a related Hubbard model and present details on the flavor-wave theory. As before, this includes an analysis of the symmetries of the magnon spectrum and a derivation of the flavor-wave chirality.

We start by considering an SU(3)-symmetric Hubbard model
\begin{equation}
	\HH_\mathrm{Hub} = - t_1 \sum_{\langle i\mu,j\nu \rangle} \sum_{\alpha=1}^{3} \left( c_{i\mu\alpha}^\dagger c_{j\nu\alpha} + \hc \right)
	- t_2 \sum_{i\mu} \sum_{\alpha=1}^{3} \left( c_{i\mu\alpha}^\dagger c_{i + \bm{\Delta}_\mu,\mu,\alpha} + \hc \right)
	+ U \sum_{i\mu} \sum_{\alpha<\beta} c_{i\mu \alpha}^\dagger c_{i\mu\alpha} c_{i\mu \beta}^\dagger c_{i\mu\beta},
	\label{eq:HHub}
\end{equation}
in which fermions of three different flavors hop between the sites $(i\mu)$ of a hexatriangular lattice, see Fig.~1(b) in the main text, while experiencing an on-site repulsion $U>0$. The amplitudes $t_1$ and $t_2$ apply, respectively, to the hopping between nearest-neighbor sites and second-neighbor sites connected by dashed lines in Fig.~1(b). Accordingly, by setting the lattice spacing to one and choosing a coordinate system whose $x$ axis is parallel to a nearest-neighbor bond, we can write $\bm{\Delta}_1 = (0,\sqrt{3})$, $\bm{\Delta}_2 = (3,\sqrt{3})/2$, and $\bm{\Delta}_3 = (-3,\sqrt{3})/2$. Models such as Eq.~\eqref{eq:HHub} can be directly implemented in cold-atom settings, which also allow one to tune the parameters $t_1$, $t_2$, and $U$ \cite{knap23,gorshkov10}.

At an average filling of one particle per site (i.e., $1/3$ filling), the low-energy behavior of $\HH_\mathrm{Hub}$ in the strong-coupling limit $U \gg t_1, t_2$ is described by the Heisenberg Hamiltonian \cite{toth10}
\begin{equation}
	\HH_S = \frac{J_{1}}{3} \sum_{\langle i\mu,j\nu \rangle} S_{i\mu}^{\beta\alpha}S_{j\nu}^{\alpha\beta}
	+ \frac{J_{2}}{3} \sum_{i\mu} S_{i\mu}^{\beta\alpha} S_{i+\bm{\Delta}_\mu, \mu}^{\alpha\beta},
	\label{eq:H SU3}
\end{equation}
with antiferromagnetic couplings $J_n \propto t_n^2/U > 0$, for $n=1,2$. The generators $S_{i\mu}^{\beta\alpha} = \Ket{\beta_{i\mu}} \Bra{\alpha_{i\mu}}$ at a given site $(i\mu)$ form a fundamental representation of SU(3), and the bilinears $S_{i\mu}^{\beta\alpha} S_{j\nu}^{\alpha\beta} = \Ket{\beta_{i\mu} \alpha_{j\nu}} \Bra{\alpha_{i\mu} \beta_{j\nu}}$ have the effect of swapping the states at sites $(i\mu)$ and $(j\nu)$ \cite{toth10,bauer12}.

We note that the $S^{\alpha\beta}$ are not the $N^2-1=8$ independent generators given, in the fundamental representation, by one-half of the Gell-Mann matrices $S^n = \lambda_{n}/2$ \cite{zee_book}. If we denote the identity matrix as $\mathds{1}$ and identify the indices $\left\{ A,B,C \right\} \equiv \left\{ 1,2,3 \right\}$, the relation between the two sets of generators can be expressed as
\begin{align}
	S^{AA} &= \frac{M}{3} \mathds{1} + S^3 + \frac{1}{\sqrt{3}} S^8,
	&
	S^{AB} &= S^1 + iS^2,
	&
	S^{BA} &= S^1 - iS^2,
	\nonumber \\
	S^{BB} &= \frac{M}{3} \mathds{1} - S^3 + \frac{1}{\sqrt{3}} S^8
	&
	S^{AC} &= S^4 + iS^5,
	&
	S^{CA} &= S^4 - iS^5,
	\nonumber \\
	S^{CC} &= \frac{M}{3} \mathds{1} - \frac{2}{\sqrt{3}} S^8,
	&
	S^{BC} &= S^6 + iS^7,
	&
	S^{CB} &= S^6 - iS^7,
	\label{eq:generatormap}
\end{align}
with $M=1$. Using properties of the Gell-Mann matrices, it is straightforward to verify that Eq.~\eqref{eq:generatormap} yields the commutation relations $\comm{S^{\alpha\beta}}{S^{\mu\nu}} = \delta_{\beta\mu} S^{\alpha\nu} - \delta_{\nu\alpha} S^{\mu\beta}$ and the condition $\sum_{\alpha} S^{\alpha\alpha} = M \mathds{1}$ quoted in the main text.


\subsection{Flavor-wave theory}
\label{subsec:lfwt}

\begin{figure}
\includegraphics[height=6cm]{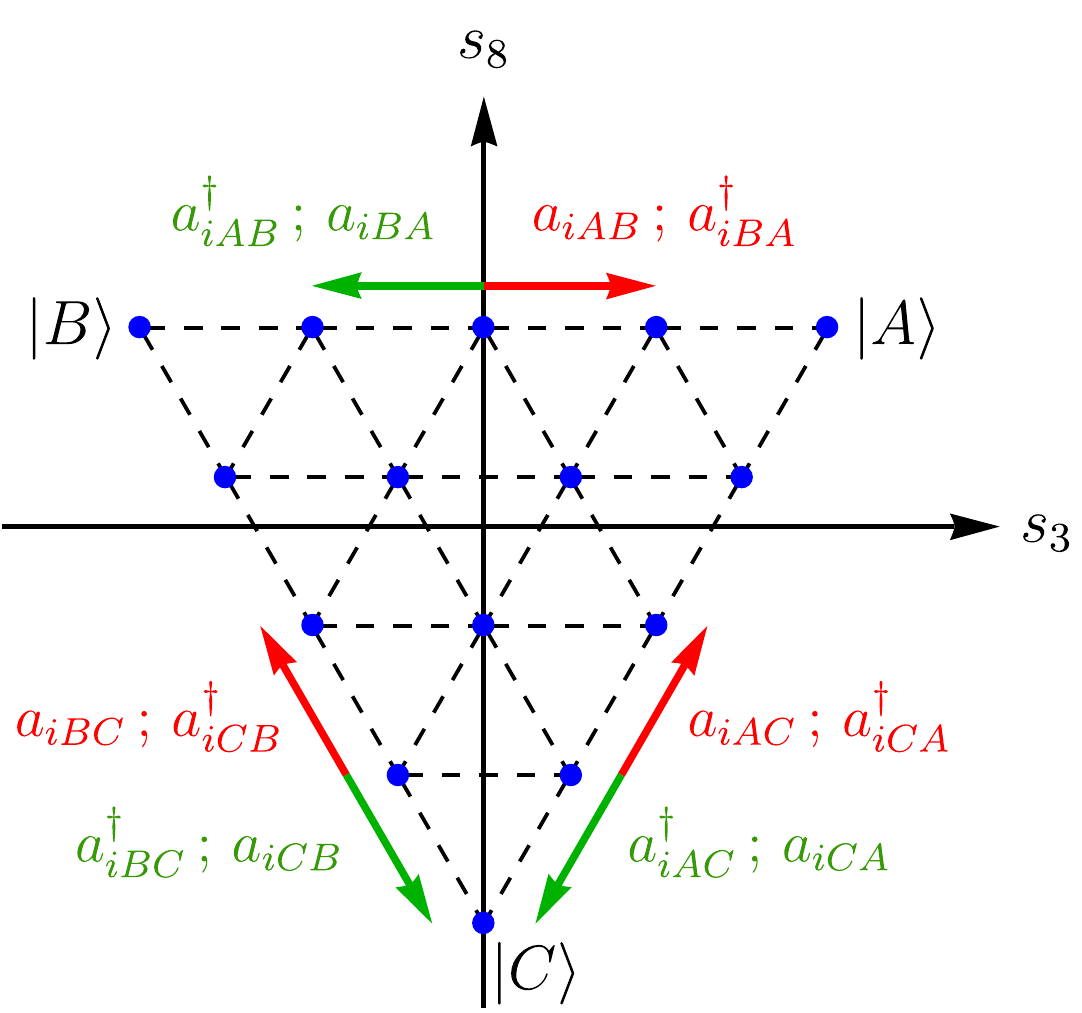}
\caption{Weight diagram of the $\left\{4,0\right\}$ totally symmetric irreducible representation of the $\mathrm{SU}\left(3\right)$ algebra. $s_3$ and $s_8$ correspond to the eigenvalues of the independent diagonal generators of $\mathrm{SU}\left(3\right)$, which can be expressed as $S^3 = \left(S^{AA}-S^{BB}\right)/2$ and $S^{8}=\sqrt{3}\left(S^{AA}+S^{BB}\right)/2-M/\sqrt{3}$, respectively. The maximum weight state $\Ket{A}$ corresponds to a fully polarized moment along the $A$ direction, and its Weyl reflections yield the two complementary states $\Ket{B}$ and $\Ket{C}$. The action of the different HP bosons is indicated by the outer arrows.}
\label{fig:weights}
\end{figure}

For sufficiently small $J_2/J_1$, the ground state of Eq.~\eqref{eq:H SU3} is a three-sublattice, three-color state \cite{laeuchli06,tsunetsugu06,bauer12}, which is adiabatically connected to the product state $\ket{\psi_0}$ in Eq.~(2) and can be thought of as the SU(3)-analogue of a collinear Néel state.
To study the excitations that arise on top of the ground state, we first generalize $\HH_S$ so that each local moment transforms under the $\left\{ M,0 \right\}$ irreducible representation of SU(3), which is furnished by completely symmetric tensors with $M$ indices and has dimension $\left(M+1\right)\left(M+2\right)/2$ \cite{zee_book}. For a given $M$, the Hilbert space of a  single spin is spanned by the (common) eigenvectors of the commuting generators $S^3$ and $S^8$. Each element of this eigenbasis has a unique set of eigenvalues $(s_3, s_8)$, and can thus be identified as a point in a ``weight diagram'' such as the one depicted in Fig.~\ref{fig:weights}.

Similarly to spin-wave theory, flavor-wave theory describes fluctuations around a reference state via an expansion in powers of $1/\sqrt{M}$. Here, $\ket{\psi_0}$ happens to be the appropriate reference state, since it is the product state that minimizes $\left\langle \HH \right\rangle$ \cite{bauer12}. For indices $\alpha,\beta \ne \mu$, we can employ the generalized HP representation
\begin{align}
S_{i\mu}^{\mu\mu} & =M-\sum_{\beta\ne\mu}a_{i\mu\beta}^{\dagger}a_{i\mu\beta}, 
& 
S_{i\mu}^{\mu\alpha} & =\sqrt{M-\sum_{\beta\ne\mu}a_{i\mu\beta}^{\dagger}a_{i\mu\beta}}\;a_{i\mu\alpha}\approx\sqrt{M}a_{i\mu\alpha},
\nonumber \\
S_{i\mu}^{\alpha\beta} & =a_{i\mu\alpha}^{\dagger}a_{i\mu\beta}, & S_{i\mu}^{\alpha\mu} & =a_{i\mu\alpha}^{\dagger}\sqrt{M-\sum_{\beta\ne\mu}a_{i\mu\beta}^{\dagger}a_{i\mu\beta}}\approx\sqrt{M}a_{i\mu\alpha}^{\dagger}.
\label{eq:HP SU3}
\end{align}
In terms of Fig.~\ref{fig:weights}, the effect of the creation of a boson $a_{i\mu\beta}^{\dagger}$ can be visualized as follows. Let $\overrightarrow{\mu\beta}$ be the vector that connects the points corresponding to the states $\ket{\mu}$ and $\ket{\beta}$. If the spin at site $(i\mu)$ is in an eigenstate $\ket{s_3, s_8}$ of $S^3$ and $S^8$, then $a_{i\mu\beta}^{\dagger} \ket{s_3, s_8}$ will either (a) be proportional to the first state $\ket{\tilde{s}_3, \tilde{s}_8}$ reached when moving away from $(s_3,s_8)$ along $\overrightarrow{\mu\beta}$ or (b) vanish, if that state does not exist in the weight diagram of the chosen irreducible representation. Formally, the shift that occurs in case (a) corresponds to a translation by one of the $N(N-1) = 6$ roots $\bm{\alpha}$ of the SU(3) algebra: $(s_3,s_8) \mapsto (s_3,s_8) + \bm{\alpha}$. Given that there are two bosonic species per sublattice, the system will have six magnon bands in total.

If we expand the spin bilinear appearing in the $J_{2}$ term of Eq.~\eqref{eq:HP SU3} up to $\mathcal{O} \left(M^0\right)$, we find
\begin{align}
	\sum_{\alpha\beta} S_{i\mu}^{\beta\alpha} S_{j\mu}^{\alpha\beta} &= S_{i\mu}^{\mu\mu}S_{j\mu}^{\mu\mu}+\sum_{\alpha\ne\mu}\left(S_{i\mu}^{\mu\alpha}S_{j\mu}^{\alpha\mu}+S_{i\mu}^{\alpha\mu}S_{j\mu}^{\mu\alpha}\right)+\mathcal{O}\left(M^{0}\right)\nonumber \\
	&= M^{2} 
	+ M\sum_{\alpha\ne\mu} \left(a_{i\mu\alpha}a_{j\mu\alpha}^{\dagger}+a_{i\mu\alpha}^\dagger a_{j\mu\alpha} - a_{i\mu\alpha}^\dagger a_{i\mu\alpha} - a_{j\mu\alpha}^\dagger a_{j\mu\alpha} \right)
	+ \mathcal{O}\left(M^{0}\right).
	\label{eq:su3J2bilinear}
\end{align}
Assuming that $\nu\ne\mu$, we can write the product in the $J_{1}$ term up to the same order:
\begin{align}
	\sum_{\alpha\beta} S_{i\mu}^{\beta\alpha}S_{j\nu}^{\alpha\beta} 
	&= S_{i\mu}^{\mu\mu} S_{j\nu}^{\mu\mu} + S_{i\mu}^{\nu\nu} S_{j\nu}^{\nu\nu} 
	+ \left[ S_{i\mu}^{\mu\nu} S_{j\nu}^{\nu\mu}
	+ \sum_{\alpha \ne \mu,\nu} \left( S_{i\mu}^{\mu\alpha} S_{j\nu}^{\alpha\mu} + S_{i\mu}^{\nu\alpha} S_{j\nu}^{\alpha\nu} \right) + \hc \right]
	+ \mathcal{O} \left(M^{0}\right)
	\notag \\
	&= M \left( a_{j\nu\mu}^\dagger a_{j\nu\mu} + a_{i\mu\nu}^\dagger a_{i\mu\nu} + a_{i\mu\nu} a_{j\nu\mu} + a_{i\mu\nu}^\dagger a_{j\nu\mu}^\dagger \right)
	\notag \\
	&+ \sqrt{M} \sum_{\alpha \ne \mu,\nu} \left( a_{i\mu\alpha} a_{j\nu\alpha}^\dagger a_{j\nu\mu} + a_{i\mu\alpha}^\dagger a_{i\mu\nu} a_{j\nu\alpha} + \hc \right)
	+ \mathcal{O}\left(M^{0}\right).
	\label{eq:su3J1bilinear}
\end{align}
Interestingly, we see that the $1/\sqrt{M}$ expansion in Eq.~\eqref{eq:su3J1bilinear} generates terms that are \emph{cubic} in HP bosons for any $N>2$, including the present $N=3$ case. This is a relevant discrepancy with respect to the spin-wave theory of any collinear SU(2) Heisenberg magnet and can have significant consequences for the lifetime of SU($N>2$) magnons. We will return to this point later on, after analyzing the quadratic terms.


\subsubsection{Linear flavor-wave theory}

By truncating Eqs.~\eqref{eq:su3J2bilinear} and \eqref{eq:su3J1bilinear} at $\mathcal{O} \left(M\right)$ and applying a Fourier transform for a system with $N_s = 3 N_c$ sites, we obtain the linear flavor-wave Hamiltonian
\begin{align}
	\HH_\mathrm{LFW} 
	&= N_{s} \left[ \frac{M^{2} J_{2}}{3} - M J_{1} \left(1 - 2\tilde{\eta}\right) \right]
	+ MJ_{1} \sum_{\vec{k}\mu} \sum_{\nu > \mu}
	\begin{pmatrix}
		a_{\vec{k}\mu\nu}^\dagger &	a_{-\vec{k},\nu\mu}
	\end{pmatrix}
	\begin{pmatrix}
		1 - 2\tilde{\eta} \tilde{\Gamma}_{\vec{k}\mu} & \tilde{\gamma}_{\vec{k},\mu\nu} \\
		\tilde{\gamma}_{\vec{k},\mu\nu}^* & 1 - 2\tilde{\eta} \tilde{\Gamma}_{\vec{k}\nu}
	\end{pmatrix}
	\begin{pmatrix}
		a_{\vec{k}\mu\nu} \\
		a_{-\vec{k},\nu\mu}^\dagger
	\end{pmatrix},
	\label{eq:hlfw2}
\end{align}
with $\tilde{\eta} = J_2/(3J_1)$, $\tilde{\Gamma}_{\vec{k}\mu} = 1 - \cos\left( \vec{k} \cdot \bm{\Delta}_{\mu} \right)$, and
\begin{align}
	\tilde{\gamma}_{\vec{k},\mu\nu} &= 
	\begin{cases}
		\tilde{\gamma}_\vec{k} = \frac{1}{3} \sum_{\bm{\delta}} e^{i \vec{k} \cdot \bm{\delta}} & \text{if } \nu = \mu+1,
		\\
		\tilde{\gamma}_\vec{k}^* & \text{if } \nu = \mu+2,
	\end{cases}
\end{align}
defined in terms of the nearest-neighbor vectors $\bm{\delta} \in \left\{ \frac{1}{2} \left( -1, \sqrt{3} \right) ; \frac{1}{2} \left( -1, -\sqrt{3} \right) ; \left(1,0\right)\right\}$. The quadratic part of $\HH_\mathrm{LFW}$ is identical to Eq.~(5). Moreover, as stated in the main text, the decomposition of $\HH_S^{(2)}$ into separate $2\times 2$ blocks for every momentum $\vec{k}$ is dictated by the invariance of $\HH$ and $\ket{\psi_0}$ under global transformations generated by $S_\mathrm{tot}^{\alpha\alpha}$ or, equivalently, by $S_\mathrm{tot}^3$ and $S_\mathrm{tot}^8$. Each $2\times 2$ block is associated with a pair of arrows in Fig.~\ref{fig:weights} or, in other words, with a pair of opposite roots $\pm\bm{\alpha}$ of SU(3).
For general $N$, the linear flavor-wave Hamiltonian of an SU($N$)-symmetric model breaks up into $N(N-1)/2$ two-dimensional blocks for every momentum $\vec{k}$, with each block corresponding to a different pair of roots $\pm\bm{\alpha}$ of the SU($N$) algebra.

Noting that the $2\times2$ blocks in Eq.~\eqref{eq:hlfw2} have the same structure as those in Eq.~\eqref{eq:h02}, we can use the results from Sec.~\ref{subsec:su2swt} to diagonalize the linear-flavor wave Hamiltonian here as well. Namely, after applying block-dependent Bogoliubov transformations
\begin{equation}
	\Psi_{\vec{k},\mu\nu}
	=
	\begin{pmatrix}
		a_{\vec{k}\mu\nu} \\
		a_{-\vec{k},\nu\mu}^\dagger
	\end{pmatrix}
	=
	\begin{pmatrix}
		u_{\vec{k}\mu\nu} & v_{\vec{k}\mu\nu} e^{i \varphi_{\vec{k}\mu\nu} } \\
		v_{\vec{k}\mu\nu} e^{-i \varphi_{\vec{k}\mu\nu} } & u_{\vec{k}\mu\nu}
	\end{pmatrix}
	\begin{pmatrix}
		\alpha_{\vec{k},\nu\mu} \\
		\alpha_{-\vec{k},\mu\nu}^\dagger
	\end{pmatrix},
	\label{eq:Bog SU3}
\end{equation}
we arrive at
\begin{align}
	\HH_\mathrm{LFW} &= \frac{N_s}{3} \left[ M^2 J_2 - 3M J_1 \left(1 - 2\tilde{\eta}\right) \right]
	+ \sum_{\vec{k}\mu} \sum_{\nu \ne \mu} 
	\left( \omega_{\vec{k}\mu \nu} \alpha_{\vec{k}\mu \nu}^\dagger \alpha_{\vec{k}, \mu\nu} + \frac{1}{2} \right),
	\label{eq:H postBog SU3}
\end{align}
with $\omega_{\vec{k}\mu\nu}$ given by Eq.~(6) of the main text. The subindices of the magnon operators in Eq.~\eqref{eq:Bog SU3} were chosen so that $\alpha_{\vec{k},\mu\nu}^\dagger$ has quantum numbers $\Delta S_\mathrm{tot}^{\alpha\alpha} = \delta_{\alpha\mu} - \delta_{\alpha\nu}$ (see Table~\ref{tab:DSmumu}).

\begin{table}
	\centering
	\newcolumntype{C}[1]{>{\centering \arraybackslash}m{#1}}
	\caption{Effect of the six creation operators on the conserved quantities $S_\mathrm{tot}^{\mu\mu} = \sum_j S_j^{\mu\mu}$.}
	\begin{tabular}{ *{7}{C{0.135\columnwidth}} }
		\hline
		\hline
		& $\alpha_{\vec{k}AB}^\dagger$ & $\alpha_{\vec{k}BA}^\dagger$ & $\alpha_{\vec{k}AC}^\dagger$ & $\alpha_{\vec{k}CA}^\dagger$ & $\alpha_{\vec{k}BC}^\dagger$ & $\alpha_{\vec{k}CB}^\dagger$ \\
		\hline
		$\Delta S_\mathrm{tot}^{AA}$ & $+1$ & $-1$ & $+1$ & $-1$ & 0 & 0 \\
		$\Delta S_\mathrm{tot}^{BB}$ & $-1$ & $+1$ & 0 & 0 & $+1$ & $-1$ \\
		$\Delta S_\mathrm{tot}^{CC}$ & 0 & 0 & $-1$ & $+1$ & $-1$ & $+1$ \\
		\hline	
		\hline
	\end{tabular}
	\label{tab:DSmumu}
\end{table}


\subsubsection{Cubic terms and magnon decay}

As demonstrated in Eq.~\eqref{eq:su3J1bilinear}, the $1/\sqrt{M}$ expansion about an $N$-color, $N$-sublattice state $\Ket{\psi_0}$ of an SU($N>2$) magnet yields terms that are \emph{cubic} in HP bosons. Fundamentally, the exclusion of the $N=2$ case can be understood as a consequence of the conservation of $S_\mathrm{tot}^z$. This demands that every $T=0$ fluctuation process generated on top of $\Ket{\psi_0}$ entail, to all orders in $1/\sqrt{M}$, the creation of an equal number of HP bosons on sublattices $A$ and $B$. Hence, symmetry forbids the presence of any terms with an odd number of bosons in the spin-wave Hamiltonian.

Despite having a larger number of conserved quantities, SU($N>2$) magnets are not subject to the same constraint. To exemplify this, consider the SU(3) system above and let $(iA)$ and $(jB)$ be two nearest-neighbor sites. Starting with a state $a_{iAC}^\dagger \Ket{\psi_0}$, we can envisage a cubic process $a_{iAB}^\dagger a_{jBC}^\dagger a_{iAC}$ where the initial $C$ HP boson is transferred from $(iA)$ to $(jB)$ and a second, $B$ boson is created at $(iA)$.
Using Fig.~\ref{fig:weights}, one can easily verify that the vectors representing the three bosonic operators add up to zero, thereby confirming that the process is compatible with the global SU(3) symmetry.

The presence of such anharmonic terms implies that, under the appropriate kinematic conditions, a single magnon can decay into pairs of excitations. This effect has been studied extensively in the context of SU(2) magnets with \emph{noncollinear} order \cite{zhitomirsky13}, and is known not only as a potential source of broadening in the spin-wave spectrum \cite{winter17}, but also of strong $1/S$ corrections to observables \cite{chernyshev09}. Based on this, we expect that similar effects can arise in \emph{any} SU($N$) magnet with long-range order. While the existence of cubic terms is not specific to altermagnets, the aforementioned phenomena may acquire particularly strong spatial, flavor, and chirality dependencies in the presence of SU($N$) altermagnetism, which may generate unique transport properties \cite{costa24}.


\subsection{Spin point group and symmetries of the flavor-wave spectrum}
\label{subsec:symmetries lfw}

We now turn to an analysis of the symmetries of the flavor-wave spectrum in order to rationalize the crossings and partial degeneracies observed in Fig.~2(b) of the main paper. Similarly to the SU(2) case, the fact that
\begin{equation}
	\omega_{\vec{k}\mu\nu} = \omega_{-\vec{k}, \mu\nu}
	\label{eq:sym1 su3}
\end{equation}
follows from an antiunitary symmetry $T_{0} =\ssop{B_0}{\Theta}$, where $B_0$ is a transformation (including the effects of time reversal) that leaves $\ket{\psi_0}$ invariant. 

Additional symmetries of the spectrum are associated with the nontrivial spin point group of the system, i.e., the set of elements that combine point group operations with independent transformations in spin space to leave both $\HH_S$ and $\ket{\psi_0}$ invariant. The term ``nontrivial'' here refers to the fact that, apart from the identity $\ssop{E}{E}$, we only consider transformations that affect real space. Generally, the number and types of allowed spin point groups in a given system are restricted by lattice symmetries and can be fully categorized as follows \cite{litvin74,litvin77}. One starts by identifying the point group $\mathbf{G}$ of the underlying lattice and constructing all normal subgroups $\mathbf{H}_{a}\subset\mathbf{G}$, i.e., the subgroups that fulfill $g\mathbf{H}_{a}g^{-1}=\mathbf{H}_{a},\;\forall g\in\mathbf{G}$. This leads to a series of quotient groups $\mathbf{G}/\mathbf{H}_{a}=\left\{ \mathbf{H}_{a},g_{2}\mathbf{H}_{a},\ldots,g_{n}\mathbf{H}_{a}\right\}$ of order $n=N\left(\mathbf{G}\right)/N\left(\mathbf{H}_{a}\right)$. For each $a$, one can find a subgroup $\mathbf{B}_{a}=\left\{ E,B_{2},\ldots,B_{n}\right\}$ of operations in spin space that is isomorphic to $\mathbf{G}/\mathbf{H}_{a}$. A spin point group $\mathbf{X}^{\left(a\right)}$ is then formed by pairwise composition of the elements of both subgroups:
\begin{align}
	\mathbf{X}^{\left(a\right)} & =\left[E||\mathbf{H}_{a}\right]+\left[B_{2}||g_{2}\mathbf{H}_{a}\right]+\cdots+\left[B_{n}||g_{n}\mathbf{H}_{a}\right].
\end{align}

The spatial symmetries of the hexatriangular lattice are fully described by the wallpaper group $p31m$, whose associated point group 
\begin{equation}
	\mathbf{G}=D_{3}=\left\{ E, R, R^{2}, \mathcal{M}_A, \mathcal{M}_B, \mathcal{M}_C \right\} .
\end{equation}
is composed of 120° counterclockwise rotations, $R$, about the center of a filled triangle and mirror operations, $\mathcal{M}_{\mu}$, that exchange sublattices $(\mu-1)$ and $(\mu+1)$. The normal subgroups of $\mathbf{G}$ and the resulting quotient groups are
\begin{align}
	\mathbf{H}_{1} & =\mathbf{G}, & \mathbf{G}/\mathbf{H}_{1} & =\left\{ \mathbf{H}_{1}\right\} ,
	\\
	\mathbf{H}_{2} & =\left\{ E,R,R^{2}\right\} \cong C_{3}, & \mathbf{G}/\mathbf{H}_{2} & =\left\{ \mathbf{H}_{2}, \mathcal{M}_A \mathbf{H}_{2}\right\} \cong Z_{2},
	\\
	\mathbf{H}_{3} & =\left\{ E\right\} , 
	& 
	\mathbf{G}/\mathbf{H}_{3} & =\left\{ \mathbf{H}_{3}, R\mathbf{H}_{3}, R^{2}\mathbf{H}_{3}, \mathcal{M}_A\mathbf{H}_{3}, \mathcal{M}_B\mathbf{H}_{3}, \mathcal{M}_C\mathbf{H}_{3}\right\} \cong S_{3},
\end{align}
where we have used the symbol $\cong$ to indicate isomorphisms between groups. In particular, the SU(3) altermagnet is characterized by
\begin{equation}
	\mathbf{X}^{\left(3\right)} = \ssop{E}{E} + \ssop{(ABC)}{R} + \ssop{(ACB)}{R^2}  + \ssop{(BC)}{\mathcal{M}_A} + \ssop{(AC)}{\mathcal{M}_B} + \ssop{(AB)}{\mathcal{M}_C},
	\label{eq:su3spinpointgroup}
\end{equation}
where the SU(3) transformations on the left-hand side of the double bars are represented as permutations of the states $\ket{A}$, $\ket{B}$, and $\ket{C}$ (see Fig.~\ref{fig:weights}). For instance, $\left(BC\right) = e^{-i\pi S^6}$ swaps $\ket{B}$ and $\ket{C}$ while leaving $\ket{A}$ unaffected (up to a phase).

We can now perform an analysis similar to the one from Sec.~\ref{subsec:symmetries lsw} to derive symmetries of the linear flavor-wave spectrum. Consider, for instance, the transformation $T_A = \left[\left(BC\right) || \mathcal{M}_A \right]$. This acts on the SU(3) generators as $T_{A} S_{i\mu}^{\alpha\beta} T_{A}^{-1} = S_{j\tilde{\mu}}^{\tilde{\alpha}\tilde{\beta}}$, with $(\tilde{A},\tilde{B},\tilde{C}) = (A,C,B)$ and $\vec{r}_{j\tilde{\mu}} = \mathcal{M}_A [\vec{r}_{i\mu}]$. By using Eq.~\eqref{eq:HP SU3}, we find that
\begin{align}
	T_{A} a_{i\mu\beta} T_{A}^{-1} &= a_{j\tilde{\mu}\tilde{\beta}},
	\\
	T_{A} a_{\vec{k}\mu\beta} T_{A}^{-1} &
	= \frac{1}{\sqrt{N_{c}}} \sum_{i} e^{ -i \vec{k} \cdot \vec{r}_{i\mu}} a_{j\tilde{\mu}\tilde{\beta}}
	= \frac{1}{\sqrt{N_{c}}} \sum_{j} e^{-i \mathcal{M}_{A}[\vec{k}] \cdot\vec{r}_{j\tilde{\mu}} } a_{j\tilde{\mu}\tilde{\beta}}
	= a_{\tilde{\vec{k}} \tilde{\mu} \tilde{\beta}},
\end{align}
where we have introduced the shorthand notation $\tilde{\vec{k}} = \mathcal{M}_A [\vec{k}]$. We shall now apply this transformation to the quadratic part $\HH_S^{(2)} = \sum_{\vec{k}\mu} \sum_{\nu>\mu} \Psi_{\vec{k},\mu\nu}^\dagger \mathbb{M}_{\vec{k}, \mu\nu} \Psi_{\vec{k},\mu\nu}$ of the linear flavor-wave Hamiltonian and manipulate the result so as to restore its original form. The ensuing changes to the matrices
\begin{equation}
	\mathbb{M}_{\vec{k},\mu\nu} =
	\begin{pmatrix}
		A_{\vec{k},\mu\nu} & C_{\vec{k}} \\
		C_{\vec{k}}^* & B_{\vec{k},\mu\nu}
	\end{pmatrix},
\end{equation}
whose elements can be read off from Eq.~\eqref{eq:hlfw2}, will then allow us to extract the symmetries of the spectrum. From
\begin{align}
	T_A \HH_S^{(2)} T_A^{-1} &=
	\sum_{\vec{k}\mu} \sum_{\nu>\mu} \Psi_{\tilde{\vec{k}},\tilde{\mu}\tilde{\nu}}^\dagger \mathbb{M}_{\vec{k},\mu\nu} \Psi_{\tilde{\vec{k}},\tilde{\mu}\tilde{\nu}}
	\notag \\
	&= \sum_{\vec{k}} \left[
	\Psi_{\vec{k},AB}^\dagger \mathbb{M}_{\tilde{\vec{k}},AC} \Psi_{\vec{k},AB}
	+ \Psi_{\vec{k},AC}^\dagger \mathbb{M}_{\tilde{\vec{k}},AB} \Psi_{\vec{k},AC}
	+ \Psi_{\vec{k},BC}^\dagger 
	\begin{pmatrix}
		B_{-\tilde{\vec{k}},BC} & C_{-\tilde{\vec{k}}} \\
		C_{-\tilde{\vec{k}}}^* & A_{-\tilde{\vec{k}},BC}
	\end{pmatrix}
	\Psi_{\vec{k},BC}
	\right]
	\label{eq:T2 A}
\end{align}
and Eq.~\eqref{eq:disp SU2}, we conclude that
\begin{align}
	\left( \omega_{\vec{k},AB},\omega_{\vec{k},BA} \right) &= \left( \omega_{\tilde{\vec{k}},AC},\omega_{\tilde{\vec{k}},CA}	\right)
	&
	&\text{and}
	&
	\omega_{\vec{k}, BC} = \omega_{-\tilde{\vec{k}}, CB}
	\label{eq:sym2 su3}
\end{align}
When combined with Eq.~\eqref{eq:sym1 su3}, this accounts for all band crossings Figs.~1(c,d) display along the lines $k_{x}=0$ and $k_{y}=0$, as well as for the symmetry of the bands under the mirror operations $\left(k_{x},k_{y}\right)\longmapsto\left(\pm k_{x},\mp k_{y}\right)$. Similar results follow from the symmetries $T_B = \ssop{(AC)}{\mathcal{M}_B}$ and $T_C = \ssop{(AB)}{\mathcal{M}_C}$. Together with Eqs.~\eqref{eq:sym1 su3} and \eqref{eq:sym2 su3}, these give a full account of the symmetries of the flavor-wave spectrum.


\subsection{Flavor-wave chirality}
\label{subsec:SU3 chirality}

We are now in position to compare the dynamics of the SU(3) spins in the excited states
\begin{equation}
	\Ket{\phi_{\vec{k}\lambda\lambda'}} = \sum_{n=0}^{\infty} c_{n} \Ket{n_{\vec{k}\lambda\lambda'}}
	= \sum_{n=0}^{\infty} \frac{c_{n}}{\sqrt{n!}} \left(\alpha_{\vec{k}\lambda\lambda'}^{\dagger}\right)^{n} \Ket{0},
\end{equation}
where $\left(\lambda \lambda'\right)$ is a permutation of fixed indices $\left(\mu\nu\right)$, with $\nu>\mu$, and $\Ket{0}$ denotes the ground state of the linear flavor-wave Hamiltonian \eqref{eq:H postBog SU3}. Our goal below will be to show that the states $\ket{\phi_{\vec{k}\mu\nu}}$ and $\ket{\phi_{\vec{k}\nu\mu}}$ have opposite chiralities, in a sense we shall explain shortly. Throughout this subsection, we will use the symbol $\tau \ne \mu,\nu$ to indicate the third flavor/sublattice. We will also consistently employ the shorthand notation $\expval{\cdots} = \Braket{\phi_{\vec{k}\lambda\lambda'} | \cdots | \phi_{\vec{k}\lambda\lambda'}}$.

To begin, we can use the property $\comm{S_{i\tau}^{\beta\beta'}}{\alpha_{\vec{k}\lambda\lambda'}} = \comm{S_{i\tau}^{\beta\beta'}}{\alpha_{\vec{k}\lambda\lambda'}^\dagger} = 0$, which follows from Eq. \eqref{eq:Bog SU3}, to show that none of the nine expectation values
\begin{align}
	\expval{ S_{i\tau}^{\beta\beta'} (t) } &= 
	\sum_{m,n} \frac{c_m^* c_n}{\sqrt{m!n!}} 
	\Braket{0 | S_{i\tau}^{\beta\beta'}(t) \left(\alpha_{\vec{k}\lambda\lambda'}\right)^m \left(\alpha_{\vec{k}\lambda\lambda'}^\dagger\right)^n | 0}
	= \Braket{0 | S_{i\tau}^{\beta\beta'}(t) | 0} \sum_{n=0}^\infty \left|c_n\right|^2
	\nonumber \\
	&= \delta_{\beta\beta'} \left[
	\delta_{\beta\tau} \Braket{0 | M - \sum_{\gamma \ne \tau} a_{i\tau\gamma}^\dagger (t) a_{i\tau\gamma} (t) | 0}
	+ \left(1 - \delta_{\beta\tau}\right) \Braket{0 |  a_{i\tau\beta}^\dagger (t) a_{i\tau\beta} (t) | 0} \right]
	\nonumber \\
	& =\delta_{\beta\beta'} \left[\delta_{\beta\tau} \left(M - 2m_\perp\right) + \left(1 - \delta_{\beta\tau}\right) m_\perp\right].
	\label{eq:su3Sitau}
\end{align}
depend explicitly on time. This result is related to the block-diagonal structure of the linear flavor-wave Hamiltonian and reflects that the fact that $\alpha_{\vec{k}\lambda\lambda'}^\dagger$ does not affect the spins on sublattice $\tau$. In the last step of Eq.~\eqref{eq:su3Sitau}, we identified $m_{\perp} = \expval{S_{i\tau}^{\gamma\gamma}}{0} = \expval{a_{i\tau\gamma}^\dagger a_{i\tau\gamma}}{0}$ as the fluctuation-induced magnetization along the complementary flavors $\gamma\ne\tau$. By symmetry, $m_\perp$ will be the same for both $\gamma$.

With this first step, we have confirmed that the dynamics of $\Ket{\phi_{\vec{k}\lambda\lambda'}}$ is restricted to sublattices $\mu$ and $\nu$. Similarly to the SU(2) case, we can arbitrarily select one of these sublattices, say $\mu < \nu$, and compare the time evolution of $\expval{ S_{i\mu}^{\beta\beta'}(t) }$ for $(\lambda\lambda') = (\mu\nu)$ and $(\lambda\lambda') = (\nu\mu)$. We will show that the nontrivial time dependence arises \emph{exclusively} from combinations of $\beta$ and $\beta'$ included in
\begin{align}
	s_{i\mu\nu}^x & =\frac{S_{i\mu}^{\mu\nu} + S_{i\mu}^{\nu\mu}}{2}, 
	& 
	s_{i\mu\nu}^y & =\frac{S_{i\mu}^{\mu\nu} - S_{i\mu}^{\nu\mu}}{2i}, 
	& 
	s_{i\mu\nu}^z & =\frac{S_{i\mu}^{\mu\mu} - S_{i\mu}^{\nu\nu}}{2}.
	\label{eq:su2 subalgebra}
\end{align}
By using the commutation relations of SU(3) generators, one can verify that the three operators above satisfy $\comm{s_{i\mu\nu}^\alpha}{s_{i\mu\nu}^\beta} = i \eps_{\alpha\beta\gamma} s_{i\mu\nu}^\gamma$, and hence form an SU(2) subalgebra of SU(3). The large-$M$, three-color state $\Ket{\psi_0}$ then plays a role similar to the highest-weight state $\Ket{S,S}$ of an irreducible representation of an SU(2) algebra, with $S=M/2$, since $s_{i\mu\nu}^z \Ket{\psi_0} = M/2 \Ket{\psi_0}$ and $\left(s_{i\mu\nu}^x + i s_{i\mu\nu}^y \right) \Ket{\psi_0} = 0$.
Far from being accidental, these properties are a direct consequence of the fact that any SU($N>2$) algebra has $N(N-1)/2$ SU(2) subalgebras \cite{zee_book}. Therefore, the structure uncovered here will be valid for a general SU($N$) magnet with an $N$-color, $N$-sublattice ground state.

To prove our claim, we can separate the expectation values $\expval{ S_{i\mu}^{\beta\beta'}(t) }$ into two groups. The first is composed of elements with at least one upper index $\tau \ne \mu,\nu$:
\begin{align}
	\expval{ S_{i\mu}^{\tau\nu} (t) } &= \expval{ S_{i\mu}^{\nu\tau} (t) } = \expval{ a_{i\mu\nu}^{\dagger} (t) a_{i\mu\tau} (t) } = 0,
	&
	\expval{ S_{i\mu}^{\tau\tau} (t) } &= 
	\expval{ a_{i\mu\tau}^{\dagger} (t) a_{i\mu\tau} (t) } = m_{\perp},
	\notag \\
	\expval{ S_{i\mu}^{\tau\mu} (t) } &= \expval{ S_{i\mu}^{\mu\tau} (t) } \approx \sqrt{M} \expval{ a_{i\mu\tau} (t) } = 0.
\end{align}
Clearly, all of these matrix elements are independent of time. The second group contains the remaining expectation values with $\beta,\beta' \ne \tau$. Although these are four in total, the constraint $\sum_\alpha S_{i\mu}^{\alpha\alpha} = M \mathds{1}$ implies that $\expval{ S_{i\mu}^{\mu\mu}(t) } + \expval{ S_{i\mu}^{\nu\nu}(t) } = M - m_\perp$. We can thus obtain a complete set of results by computing
\begin{align}
	\expval{s_{i\mu\nu}^{+} (t)} & 
	\equiv \expval{S_{i\mu}^{\mu\nu} (t)}
	\approx \sqrt{M} \sum_{m,n} c_{m}^{*}c_{n}
	\Braket{m_{\vec{k}\lambda\lambda'} | a_{i\mu\nu} (t) | n_{\vec{k}\lambda\lambda'}}
	\nonumber \\
	&= \sqrt{\frac{M}{N_{c}}} \sum_{m,n} c_{m}^{*}c_{n} \sum_{\vec{q}} e^{i\vec{q}\cdot\vec{r}_{i\mu}}
	\Braket{m_{\vec{k}\lambda\lambda'} | u_{\vec{q}\mu\nu} \alpha_{\vec{q}\nu\mu} (t) + v_{\vec{q}\mu\nu} e^{i\varphi_{\vec{q}\mu\nu}} \alpha_{-\vec{q},\mu\nu}^{\dagger} (t) | n_{\vec{k}\lambda\lambda'}}
	\nonumber \\
	&= \sqrt{\frac{M}{N_{c}}} \left|\xi\right| e^{- i \sgn(\lambda' - \lambda) \left(\vec{k}\cdot\vec{r}_{i\mu} - \omega_{\vec{k}\lambda\lambda'}t + \arg\xi\right)}
	\left( \delta_{\lambda\nu} \delta_{\lambda'\mu} u_{\vec{k}\mu\nu} + \delta_{\lambda\mu} \delta_{\lambda'\nu} v_{\vec{k}\mu\nu} e^{i\varphi_{\vec{k}\mu\nu}} \right)
\end{align}
and
\begin{align}
	\expval{ s_{i\mu\nu}^{z} (t) } 
	&= \frac{M}{2} - \expval{a_{i\mu\beta}^{\dagger} a_{i\mu\beta}}{0}
	-\frac{2}{N_{c}} \sum_{\vec{pq}} e^{i \left(\vec{p}-\vec{q}\right) \cdot \vec{r}_{i\mu}} 
	\expval{ a_{\vec{q}\mu\nu}^{\dagger} (t) a_{\vec{p}\mu\nu} (t) }
	\nonumber \\
	&\approx \frac{M}{2} - 3m_{\perp} - \frac{2\zeta}{N_{c}} \left( \delta_{\lambda\nu} \delta_{\lambda'\mu} u_{\vec{k}\mu\nu}^2 + \delta_{\lambda\mu} \delta_{\lambda'\nu} v_{\vec{k}\mu\nu}^2 \right),
\end{align}
where $\xi$ and $\zeta$ are defined as in the SU(2) case [see Eqs.~\eqref{eq:expval S+ SU2} and \eqref{eq:m SU2}]. Given that these results have the same form as Eqs.~\eqref{eq:expval S+ SU2} and \eqref{eq:expval Sz SU2}, we conclude that the semiclassical dynamics of $S_{i\mu}^{\beta\beta'}$ in the excited state $\ket{\phi_{\vec{k}\lambda\lambda'}}$ is mapped onto the precession of a vector $\expval{\vec{s}_{i\mu\nu} (t)} = \left( \expval{s_{i\mu\nu}^{x} (t)}, \expval{s_{i\mu\nu}^{y} (t)}, \expval{s_{i\mu\nu}^{z} (t)} \right)$ around the $\unitvec{z}$ axis defined by its third component.

Based on the previous results, we can define the chirality of the flavor-wave modes in similar fashion to Eq.~\eqref{eq:su2kappa}:
\begin{equation}
	\kappa_{\lambda\lambda'} \equiv \sgn\left\{
	\left[ \expval{ \vec{s}_{i\mu\nu}(t) } \times \frac{d}{dt} \expval{ \vec{s}_{i\mu\nu}(t) }\right] \cdot \unitvec{z}
	\right\} ,
\end{equation}
As before, $(\lambda\lambda')$ represents a permutation of the indices $(\mu\nu)$ with $\mu<\nu$. The choice of $(i\mu)$ as a reference site for each pair of modes with dispersions $\omega_{\vec{k}\mu\nu}$ and $\omega_{\vec{k}\nu\mu}$ is of course arbitrary, and an equally valid definition would be to probe the dynamics from a site on sublattice $\nu$. However, the sole effect of this modification is to invert the chirality of every mode. This clarifies that the chirality index of a single mode does not carry physical significance on its own; only differences in chiralities are meaningful. A calculation analogous to the one sketched after Eq. \eqref{eq:su2kappa} yields
\begin{equation}
	\kappa_{\lambda\lambda'} = \sgn \left( \lambda' - \lambda \right)
	\qquad\text{(if \ensuremath{\xi\ne0})}.
	\label{eq:su3kappafinal}
\end{equation}
In other words, besides being related to opposite sets of quantum numbers $\Delta S_\mathrm{tot}^{\beta\beta}$, the modes with dispersions $\omega_{\vec{k},\mu\nu}$ and $\omega_{\vec{k},\nu\mu}$ also have opposite chiralities. The addition of a coupling $J_{2}\ne0$ renders these modes nondegenerate in the resulting altermagnetic phase, as in the SU(2) case. We stress, however, that the characterization of the magnon bands in terms of different sets of quantum numbers $\Delta S_\mathrm{tot}^{\beta\beta}$ offers not only a more fundamental, but also a more precise, definition of SU($N$) altermagnetism.


\section{Altermagnetic Kondo-lattice models and flavor-split bands}
\label{sec:metals}

This section contains details on the electronic altermagnetic models discussed in the main text. Besides specializing thethe Hamiltonian in Eq.~(9) to the cases $N=2$ and $3$, we construct the mean-field Hamiltonians whose diagonalization yields the electronic spectra shown in Fig.~3 of the main text.


\subsection{SU(2) checkerboard Kondo-lattice model}
\label{subsec:SU2met}

For SU(2), the Kondo-lattice model given by Eq.~(9) takes the familiar form
\begin{equation}
\HH_e = -\sum_{i\mu,j\nu} \left(t_{i\mu,j\nu} c_{i\mu\alpha}^{\dagger}c_{j\nu\alpha}+\mathrm{h.c.}\right)
- \frac{K}{4S} \sum_{i\mu} \vec{S}_{i\mu}\cdot\left(c_{i\mu\alpha}^{\dagger}\boldsymbol{\sigma}_{\alpha\beta}c_{i\mu\beta}\right),
\label{eq:HK su2}
\end{equation}
where $c$ electrons with spin $\alpha\in\left\{ \uparrow,\downarrow\right\} \equiv\left\{ 1,2\right\} $ coexist with local moments $\vec{S}_{i\mu}$ that occupy the sites of a checkerboard lattice. The hopping amplitudes are defined such that $t_{i\mu,j\nu}=t_{1}$ ($t_2$) on the solid (dashed) bonds of the lattice shown in Fig.~1(a) of the main paper. Furthermore, $\bm{\sigma} = \left(\sigma^{x},\sigma^{y},\sigma^{z}\right)$ is a vector of Pauli matrices.

After replacing the operators $\vec{S}_{i\mu}$ by their classical expectation values $\expval{\vec{S}_{i\mu}}{\psi_0} = \left(-1\right)^{\mu+1} S \unitvec{z}$, we obtain a purely electronic Hamiltonian which describes the motion of electrons in a staggered magnetic field. This mean-field Hamiltonian is spin-diagonal and can be written as
\begin{equation}
\HH_e^\mathrm{MF} = \sum_{\vec{k}\alpha}
\begin{pmatrix}
	c_{\vec{k}A\alpha}^{\dagger} &
	c_{\vec{k}B\alpha}^{\dagger}
\end{pmatrix}
\begin{pmatrix}
	-2t_{2} \cos\left(k_x + k_y\right) + \frac{\left(-1\right)^{\alpha}}{4} K & -4t_{1}\gamma_{\vec{k}}\\
	-4t_{1}\gamma_{\vec{k}} & -2t_{2} \cos\left(k_x - k_y\right) - \frac{\left(-1\right)^{\alpha}}{4} K
\end{pmatrix}
\begin{pmatrix}
	c_{\vec{k}A\alpha} \\
	c_{\vec{k}B\alpha}
\end{pmatrix}.
\end{equation}
A straightforward diagonalization of the $2\times2$ matrices above yields Eq.~(10) of the main text.


\subsection{SU(3) hexatriangular Kondo-lattice model}

As noted in the main text, the SU(3)-symmetric generalization of Eq. \eqref{eq:HK su2} is given by \cite{parcollet98}
\begin{equation}
\HH_e = -\sum_{i\mu,j\nu} \left(t_{i\mu,j\nu}c_{i\mu\alpha}^{\dagger}c_{j\nu\alpha}+\mathrm{h.c.}\right)
- \frac{K}{2M} \sum_{i\mu}\sum_{n=1}^{8} S_{i\mu}^{n}\left(c_{i\mu\alpha}^{\dagger}\lambda^{n}_{\alpha\beta}c_{i\mu\beta}\right),
\label{eq:HK su3}
\end{equation}
where $\alpha,\beta\in\left\{ 1,2,3\right\} \equiv\left\{ A,B,C\right\}$ are flavor indices and $\lambda^{n}$, with $n=1,\ldots,8$, represent the Gell-Mann matrices. Following standard notation, we have $\lambda^3$ and $\lambda^8$ as the only two diagonal matrices of the set. Meanwhile, the $S_{i\mu}^{n}$ operators are assumed to form an $\left\{ M,0 \right\}$ irreducible representation of the SU(3) algebra and are related to the nine operators $S_{i\mu}^{\alpha\beta}$ introduced via Eq.~\eqref{eq:generatormap}. The hopping amplitudes $t_{i\mu,j\nu}$ are set to $t_1$ for nearest-neighbor sites and to $t_2$ if $(i\mu)$ and $(j\nu)$ are next-nearest neighbors connected by a dashed line in Fig.~1(b) of the main text.

Similarly to Sec.~\ref{subsec:SU2met}, we now study the Hamiltonian \eqref{eq:HK su3} within a semiclassical approximation that replaces the operators $S_{i\mu}^{n}$ by their expectation value with respect to the $M\to \infty$ ground state $\Ket{\psi_0}$. Explicitly,
\begin{equation}
S_{i\mu}^{n} \mapsto \expval{S_{i\mu}^{n}}{\psi_0} = \expval{S_{i\mu}^{n}}{\mu} = \frac{M}{2} \left( \delta_{n3} + \delta_{n8} \right) \lambda^{n}_{\mu\mu}
\end{equation}
leads to the purely electronic mean-field Hamiltonian
\begin{align}
	\HH_e^{\mathrm{MF}} &= 
	-t_{1} \sum_{\langle i\mu, j\nu \rangle} c_{i\mu\alpha}^{\dagger} c_{j\nu\alpha}
	-t_{2} \sum_{i\mu} c_{i\mu\alpha}^\dagger c_{i+\bm{\Delta}_\mu,\,\mu\alpha}
	- \frac{K}{4} \sum_{j\mu} \left(\lambda^{3}_{\mu\mu} \lambda^{3}_{\alpha\beta} + \lambda^{8}_{\mu\mu} \lambda^{8}_{\alpha\beta}\right) c_{j\mu\alpha}^{\dagger} c_{j\mu\beta}
	\nonumber \\
	&= -t_{1} \sum_{\vec{k}\mu} \left( \tilde{\gamma}_\vec{k} c_{\vec{k}\mu\alpha}^{\dagger} c_{\vec{k},\mu+1,\alpha} + \hc \right)
	- \sum_{\vec{k}\mu} 
	\left[ 2t_2 \cos\left( \vec{k} \cdot \bm{\Delta}_{\mu}\right) + K \Lambda_{\mu\alpha} \right] c_{\vec{k}\mu\alpha}^{\dagger}c_{\vec{k}\mu\alpha}.
	\label{eq:HKMF su3}
\end{align}
Here, the second-neighbor vectors $\bm{\Delta}_\mu$ and the form factor $\tilde{\gamma}_\vec{k}$ are defined as in Sec.~\ref{sec:SU3ins}, whereas
\begin{equation}
\Lambda=\frac{1}{6}
\begin{pmatrix}2 & -1 & -1\\
-1 & 2 & -1\\
-1 & -1 & 2
\end{pmatrix}.
\end{equation}
Equation \eqref{eq:HKMF su3} is diagonal in the flavor index $\alpha$ and can be written in the form $\HH_e^\mathrm{MF}  = \sum_{\vec{k}\alpha} \Psi_{\vec{k}\alpha}^\dagger \mathbb{M}_{\vec{k}\alpha} \Psi_{\vec{k}\alpha}$, with $\Psi_{\vec{k}\alpha}=\begin{pmatrix}c_{\vec{k}A\alpha} & c_{\vec{k}B\alpha} & c_{\vec{k}C\alpha}\end{pmatrix}^{\top}$and
\begin{equation}
	\mathbb{M}_{\vec{k}\alpha} = -
	\begin{pmatrix}
		2t_2 \cos\left(\vec{k} \cdot \bm{\Delta}_{A}\right) + K \Lambda_{A\alpha} & 3t_1 \tilde{\gamma}_\vec{k} & 3t_1 \tilde{\gamma}^{*}_\vec{k} \\
		3t_1 \tilde{\gamma}^{*}_\vec{k} & 2t_2 \cos\left(\vec{k}\cdot\bm{\Delta}_{B}\right) + K\Lambda_{B\alpha} & 3t_1 \tilde{\gamma}_\vec{k} \\
		3t_1 \tilde{\gamma}_\vec{k} & 3t_1 \tilde{\gamma}^{*}_\vec{k} & 2t_2 \cos\left(\vec{k}\cdot\bm{\Delta}_{C}\right) + K\Lambda_{C\alpha}
	\end{pmatrix}.
\end{equation}
The electronic spectrum is then obtained by diagonalizing the matrices $\mathbb{M}_{\vec{k}\alpha}$ numerically. This yields three bands of each flavor, as shown in Fig.~3(c) of the main paper.


\subsection{Symmetries of the SU(3) electronic bands}

If $s_{i\mu}^{n}=c_{i\mu\alpha}^{\dagger}\tau_{\alpha\beta}^{n}c_{i\mu\beta}$ denotes the local flavor density of $c$ fermions, then the properties
\begin{align}
	\comm{ c_{i\mu\alpha}^{\dagger}c_{j\nu\alpha} }{ s_{i\mu}^{n} + s_{j\nu}^{n} } &= 0,
	& 
	\comm{ \sum_{n=1}^8 S_{i\mu}^n s_{i\mu}^n }{ S_{i\mu}^{n} + s_{i\mu}^n } &= 0,
\end{align}
imply that the Hamiltonian \eqref{eq:HK su3} is invariant under global SU(3) transformations acting identically on $S$ and $s$ operators:
\begin{equation}
	\comm{\HH_e}{ \prod_{j\mu}e ^{-i \left(S_{j\mu}^n + s_{j\mu}^n \right) \theta} } = 0
	\qquad \text{for } n=1,2,\ldots,8.
\end{equation}
If we additionally use the fact that the $S$ degrees of freedom order in a three-color state, which is invariant under SU(3) transformations generated by $S^{3}$ and $S^{8}$, then we find that
\begin{equation}
	\comm{ \HH_e^\mathrm{MF} }{ \prod_{j\mu} e^{-i s_{j\mu}^n \theta} } = 0
	\qquad \text{for } n=3,8.
\end{equation}
This explains why the fermionic flavor is a good quantum number, as we observed in the explicit calculations above. Note, however, that $\left[\HH_e^{\mathrm{MF}},\prod_{i\mu}e^{-is_{i\mu}^{n}\theta}\right]\ne0$ for $n\notin\left\{ 3,8\right\} $, since the mean-field Hamiltonian assumes that SU(3) symmetry is spontaneously broken in the local-moment sector.

The symmetries of the band structure are again dictated by the spin point group in Eq.~\eqref{eq:su3spinpointgroup}. For instance, the symmetry $T_A = \left[ \left(BC\right) || \mathcal{M}_A \right]$ enforces the degeneracy of $B$ (green) and $C$ (red) bands along the line $k_x = 0$ and the vertical boundaries of the Brillouin zone, which are formed by momenta that satisfy $\mathcal{M}_A [\vec{k}] = \vec{k} + \vec{G}$ for a reciprocal lattice vector $\vec{G}$. Similar considerations, paired with the fact that the bands are inversion-symmetric, yield all of the crossings observed in Fig.~3(c,d) of the main text between bands of different colors.
